\documentclass[preprint,12pt]{elsarticle}
\usepackage[utf8]{inputenc}
\usepackage[english]{babel}
\usepackage{amsmath}
\usepackage{graphicx}
\usepackage{appendix}
\usepackage[a4paper, total={7in, 8in}]{geometry}
\usepackage{multirow}
\usepackage[colorlinks=true, allcolors=blue]{hyperref}

\biboptions{sort&compress}

\makeatletter
\def\ps@pprintTitle{%
   \let\@oddhead\@empty
   \let\@evenhead\@empty
   \let\@oddfoot\@empty
   \let\@evenfoot\@oddfoot
}
\makeatother

\newcommand{\eqr}[1]{equation~\eqref{#1}}
\newcommand{\secr}[1]{section~\ref{#1}}
\newcommand{\appr}[1]{\ref{#1}}
\newcommand{\figr}[1]{figure~\ref{#1}}

\date\today

\begin{document}
\begin{frontmatter}

\title{Diffusive resettlement:  irreversible urban transitions in closed systems}

\author[label1]{Bohdan Slavko\corref{cor1}}
\author[label1]{Mikhail Prokopenko}
\author[label1]{Kirill S. Glavatskiy}

\address[label1]{Centre for Complex Systems, The University of Sydney, Sydney, NSW 2006, Australia}

\cortext[cor1]{To whom correspondence should be addressed: bohdan.slavko@sydney.edu.au.}


\begin{abstract}
We propose a non-equilibrium framework for modelling the evolution of cities, which describes the intra-urban migration as an irreversible diffusive process. We validate this framework using the actual migration data for the Australian capital cities. With respect to the residential relocation, the population is shown to be composed of two distinct groups, exhibiting different relocation frequencies. In the context of the developed framework, these groups can be interpreted as two components of a binary fluid mixture, each with its own diffusive relaxation time. Using this approach, we obtain long-term predictions of the cities' spatial structure, which defines their equilibrium population distribution.
\end{abstract}


\end{frontmatter}



\section{Introduction}
Modern cities are diverse in their spatial structure: some cities are monocentric, with a single center of business, retail and other types of activity, while some exhibit polycentric patterns in which multiple activity clusters are distributed across space~\cite{Giuliano1991,Mcmillen2001,Tsai2005,Green2007,Meijers2008}. It is well known that a city structure affects economic productivity, environmental conditions and other aspects of human life~\cite{Newman1989,Ewing2015,Li2018,Kwon2018,Li2019}. Importantly, spatial structures of cities change over time in intricate ways, with the process of intra-urban migration being one of the main drivers of the city evolution~\cite{Batty2006,Batty2013,Schneider2013,Arcaute2015,Barthelemy2016book,Arcaute2016,Barthelemy2019,Sahasranaman2019,Crosato2018,Dynamic_resettlement_paper}. Yet, there is a lack of models that can quantitatively explain and  accurately predict the city evolution in terms of intra-urban migration and the resultant patterns of an urban spatial structure.

In existing models of cities, the intra-urban migration is usually considered as a fast process in which an \textit{equilibrium} is reached very quickly \cite{Fujita1982, Harris1978, Louf2013, Crosato2018, Ellam2018, Dynamic_resettlement_paper}. Such an equilibrium is typically defined by the spatial distribution of infrastructure and employment  \cite{Fujita1982, Louf2013, Crosato2018, Wu2019, Slavko2020, Crosato2020, Slavko2020}, and {the intra-urban dynamics} typically is not  considered as an example of non-equilibrium process in the thermodynamic sense. A non-equilibrium approach is also hard to validate explicitly due to the difficulty of collecting the data. 

In this work, we consider the intra-urban migration as irreversible process, and explicitly derive the dynamics of the corresponding non-equilibrium evolution.  In doing so, we shall draw on an analogy with diffusive relaxation. This opens a way towards a systematic and coherent framework describing the human migration within cities (both temporally and spatially), as opposed to an unconstrained evolution of cities through expansion. 

Human relocation has been widely considered as diffusion in open systems \cite{Woube2005, Vahia2017}. This approach employs an apparent and direct analogy between human migration and molecular diffusion, and has shown good predictions at different scales \cite{Barthelemy2019, Barbosa2018, Balcan2009, Barthelemy2016book, Arcaute2016, Bouchaud2013}, from city growth \cite{Lenormand2015, Gonzalez2008, Louf2013} and epidemic spread \cite{Gustafson2017, Wen2018, Balcan2009} to inter-continental migration \cite{Woube2005}. These migration processes can be characterized as expansive, as they increase the area of human habitat, viewing the human society as an open system. In the modern world, however, most of the migration processes result in a redistribution of the population across already occupied locations, rather than expanding to non-occupied areas. This constraint essentially confines the migration to happen in a closed system, with a fixed area and population. In this paper we propose a diffusion model describing migration processes in a \textit{closed system} from the perspective of non-equilibrium thermodynamics. 

In general, migration from rural to urban areas, as well as expansion of metropolitan boundaries, are relatively slow and long-term processes. In contrast, the intra-urban migration happens at a much faster rate \cite{ Weidlich1990, Barthelemy2013Planning, Wu2004,Crosato2018,Dynamic_resettlement_paper,Slavko2020,Crosato2020}. Nevertheless, we argue that the intra-urban migration is a fundamental driving force shaping the long-term evolution of spatial urban structure, with external migration and spatial expansion playing only a secondary role. Thus, we aim to model urban transitions as dynamics developing in a closed system, at least in the first approximation. 

Typically, the intra-urban human mobility has been considered as a process driven by certain attractiveness of various locations within a city, perceived in terms of proximity to schools, business centres, recreational facilities, etc. \cite{Ellam2018, Crosato2018}. This notion of attractiveness was modelled both explicitly, using specific socio-economic indicators \cite{Kim2005, Perez2003}, and implicitly,  being reconstructed directly from the migration data \cite{Slavko2020}. The latter approaches can be classified as ``microscopic" \cite{Simini2012, Weidlich1988}, as they focus on a specific mechanism of human relocation. 

In this paper we propose a concise ``phenomenological" approach
which considers the migration flows from the perspective of diffusion. Importantly, we do not make any assumptions on  particular choices which may motivate individuals to relocate. Instead, we analyze their collective movement {and show that this process is similar to diffusion. As a result, we reveal the trends of population movement, analogous to the macroscopic movement of a fluid.} In doing so, we introduce a rigorous definition of an equilibrium state as the spatial configuration to which the apparent evolution relaxes, and a decomposition of the population into two distinct groups with different relocation frequencies.

This paper is organized as follows. In \secr{sec:markov} we present a general framework of irreversible evolution of a city driven by residential relocation. In \secr{sec:results} we apply this framework to the Australian Capital cities. In particular, we analyse the dynamic relocation patterns in \secr{sec:one_vs_two_components_data}, {develop the analogy between intra-urban resettlement and diffusion in \secr{sec:diffusion}, and predict} the equilibrium population distribution in \secr{sec:long_term_structure_predictions}. In \secr{sec:robustness}, we analyze {the robustness of the model}. Finally, in \secr{sec:conclusions} we summarize the findings of this work.


%
\section{{Dynamics of intra-urban migration}}\label{sec:markov}
%

We consider an urban area as a set of $N$ suburbs $i$ with a certain residential population $x_i(t)$ at time $t$. The total population at any time is fixed: $\sum_{i=1}^{N} x_i(t)=\overline{x}$. We assume that time is discrete, with the choice of the time step is  dictated by the resolution of the available data. A migration flow $T_{ij}(t)$ is defined as a change in residential location from suburb $i$ to suburb $j$. This flow is uni-directional, so that in general $T_{ij}(t) \neq T_{ji}(t)$, and the net flow $J_{ij}(t) \equiv T_{ij}(t) - T_{ji}(t) \neq 0$. Non-zero net flow indicates that the system is out-of-equilibrium. In a diffusive system, the net flow gradually decays to zero with time, as the system evolves towards an equilibrium. Such an equilibrium state is stationary on the ``macroscopic" level, showing no change in the population of each suburb. However, on the ``microscopic" level there still exists some movement of people, resulting in non-zero uni-directional flows $T_{ij}(t)$. In an equilibrium, these uni-directional flows satisfy a microscopic detailed balance, so that $T_{ij}(t) = T_{ji}(t)$, resulting in a zero net flow $J_{ij}(t)$ between each pair of suburbs.

The uni-directional flow matrix allows one to predict the future population of any suburb. In particular, the population at the next time step $t+1$ can be expressed through the migration flow $T_{ij}(t)$ at the current time step $t$ as $x_j(t+1)=\sum_{i=1}^{N}T_{ij}(t)$, where the sum includes the term $T_{jj}(t)$ accounting for immobile population. Introducing the fraction of relocated people as $p_{ij}(t) \equiv T_{ij}(t)\,/ x_i(t)$, we can write the population evolution equation as 
\begin{equation}\label{eq:forceflux1}
    X(t+1) = X(t) P(t),
\end{equation}
where $X$ is the (row) vector of the suburbs' population and $P$ is the relocation matrix denoting the fractions of relocating people between each pair of suburbs, with the diagonal elements $p_{jj}$ denoting the fraction of non-relocating residents. The column sum for each row of the relocation matrix is equal to $\sum_{j=1}^{N}p_{ij} = 1$, so $P$ is a row-stochastic matrix \cite{Grinstead2012}.

{The population evolution \eqr{eq:forceflux1} represents a simple Markov process, converging to a distinct stationary state $X_{eq}$, which we identify as the equilibrium state. In equilibrium, the population of each suburb $x_{i,eq}$ does not change in time, such that}

\begin{equation}\label{eq:stationary_total}
    X_{eq} \cdot P_{eq} = X_{eq}
\end{equation}
where $P_{eq} \equiv \lim_{t\to\infty}\,P(t)$. 
{We assume that the urban area is a closed system with no external shocks and, thus,} the relocation matrix does not change in time, i.e., $P_{eq} \approx P(0) \equiv P$.
Since matrix $P$ is row-stochastic, it has a unit eigenvalue \cite{Grinstead2012} and the vector $X_{eq}$ can be found as a left eigenvector of matrix $P$ that corresponds to the unit eigenvalue. This eigenvector is unique (up to a constant multiplier) if some power of matrix $P$ has strictly positive elements \cite{Grinstead2012}. 

Our next step is to explicitly represent the relocation dynamics in terms of both spatial and temporal {terms}. We decompose the relocation matrix according to the following structure 
\begin{equation}\label{eq:migration_matrix_decomposition}
    P = (1-\epsilon) I + \epsilon H,
\end{equation}
where $\epsilon$ is the share of people who relocate to a different suburb within a period of time. Such decomposition is known as the mover-stayer model \cite{Blumen1955} which has been used to describe relocation phenomena in biology, economics and social sciences \cite{Fuchs1988, Cook2002, Fougere2003, Frydman2004}. 

Here, matrix $H$ shows the relocation structure of those residents who moved to a different suburb (i.e., have not stayed in the same suburb). The matrix $H$ shows the spatial structure of the system and characterizes the variation in microscopic ``attractiveness" between different suburbs. {Without loss of generality, we assume $h_{ii}=0$.} Furthermore, the coefficient $\epsilon$ can be alternatively written as $\epsilon \equiv 1/\tau$, where $\tau$ is the characteristic relocation time. We will refer to it as the relocation frequency or the population mobility.

{We next extend the simple decomposition \eqref{eq:migration_matrix_decomposition}, so that the population mobility has a more complex structure than a single frequency $\epsilon$. Without loss of generality, we assume that the relocation dynamics is  governed by a discrete set of relocation frequencies $\epsilon_k$, where $k=1,2,...K$.} In the context of the population structure, this would suggest that the urban population comprises several distinct groups, which differ in their mobility. There may exist a number of classifications which differentiate population groups by their mobility, based on their ownership status (renters and home-owners), family status (singles and families), employment status (students, professionals, retirees). In this work we abstract away from the specific nature of these groups, assuming only their existence.

Expanding the structure of the population mobility does not affect the spatial structure of the relocation, i.e., the matrix $H$. {We therefore assume that the spatial structure of the relocation dynamics is the same for each population group. Although this is, in principle, a strong condition, we show in \secr{sec:heterogeneous_H} that it does not affect the results in practice.}

The equilibrium population of each component is obtained similarly to \eqr{eq:stationary_total}, as 
\begin{equation}\label{eq:stationary}
    X_{k,eq} \cdot P_{k} = X_{k,eq},
\end{equation}
We show in \appr{appendix:same_equilibrium} that the population of each component $X_{k}(t)$ converges to the equilibrium population structure
\begin{equation}\label{eq:fraction}
    X_{k,eq} = \alpha_k X_{eq},
\end{equation}
where  $\alpha_{k}$ is the total fraction of the city population belonging to the component $k$, so that $\sum_{k=1}^{C} \alpha_{k} = 1$ and $X_{eq}$ is the total equilibrium population structure which is independent of $\epsilon_k$ and $\alpha_k$. This is also illustrated in \figr{fig:convergence_example}. Thus, the total equilibrium $X_{eq}$ can be obtained using the full spatial matrix $H$, without the component-specific relocation matrices $P_k$ or even the component's fractions $\alpha_k$. This is very convenient, as the component structure of the population is not known \textit{a priori}, and in practice it is only matrix $H$ which can be obtained from the data directly.

\section{Results}\label{sec:results}
{In this section, we present the results of our framework for the Australian capital cities. First, in \secr{sec:one_vs_two_components_data}, we demonstrate that the model with homogeneous population fails to consistently describe the intra-urban migration dynamics, while a heterogeneous model resolves this issue.
Second, in \secr{sec:diffusion}, we develop an analogy between an intra-urban migration and diffusion. This allows us to interpret the heterogeneous dynamics of intra-urban migration as diffusion in a multi-component fluid mixture.
Third, in \secr{sec:long_term_structure_predictions}, we predict the equilibrium configuration of the considered cities.}

\subsection{{Revealing two-component structure of intra-urban evolution}}\label{sec:one_vs_two_components_data}

We first {analyze} the human relocation flows in eight Australian Greater Capital Areas, which represent populated metropolitan areas with diverse cultural and economic activities. The {model is} calibrated using the data from the Australian Census \cite{ABS2016}, which are reported as the migration flows $T_{ij}$ between each pair of suburbs within 1 year and 5 years, denoted as $T_{ij;1Y}$ and $T_{ij;5Y}$ respectively.  This suggests the natural choice for the time step as 1 year. The data are available for two census years, 2011 and 2016, with the migration counted backwards. {Intra-suburb migration is not considered in this analysis. The data resolution we use, Statistical Area 2, is the finest in the Australian Census for which the migration data is available.}

\begin{figure}[h!]
\centering
    \includegraphics[height=7in]{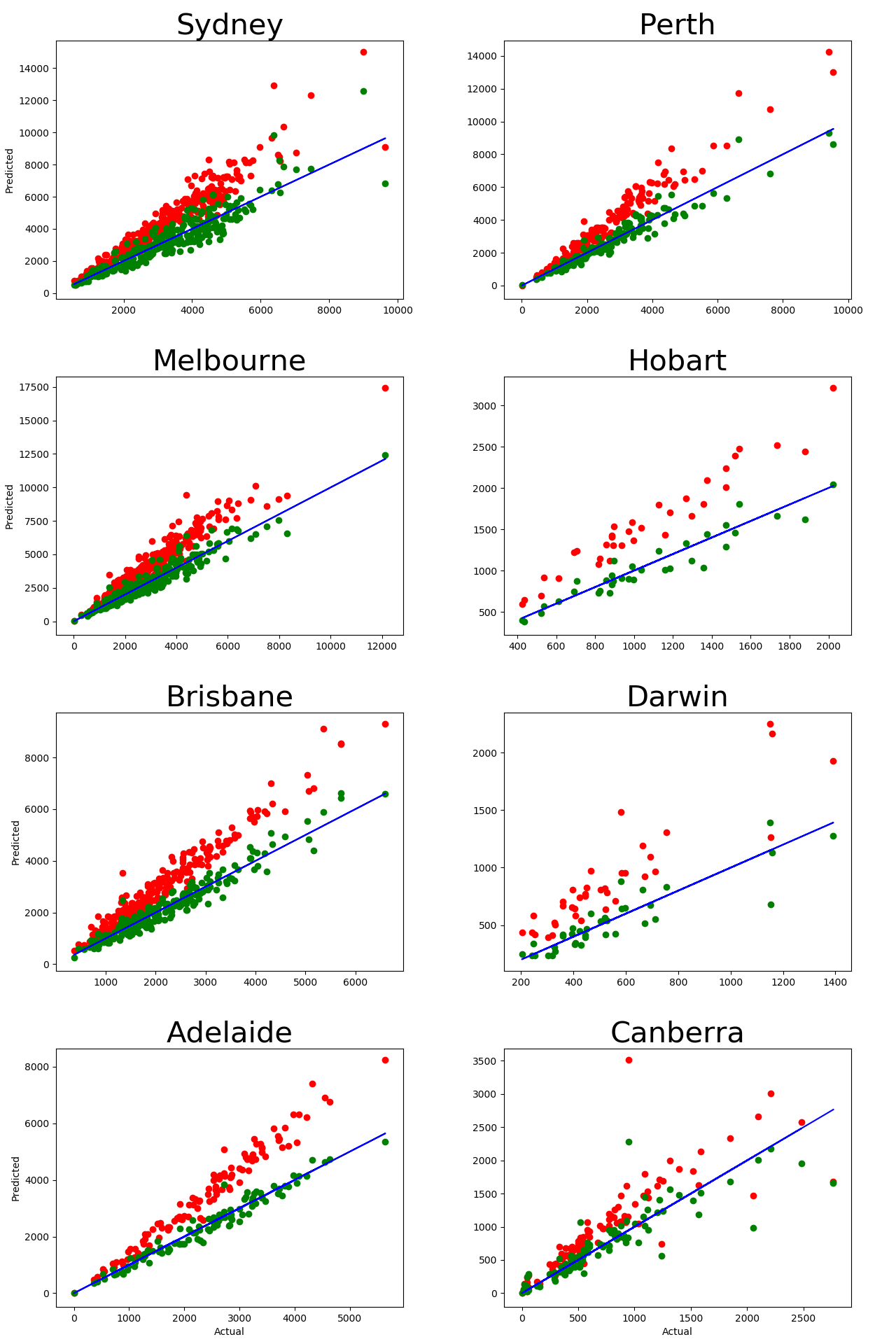}
    \caption{Number of movers in five-year migration data: actual ($\sum_{i \neq j} T_{ij;\,5Y}(2016)$) vs predicted ($\sum_{i \neq j}\hat{T}_{ij;\,5Y}(2016)$), with each dot representing one suburb. Red dots correspond to the one-component model, the green ones correspond to the two-component model. The blue solid line has the slope of $1$, showing the ideal prediction. {The corresponding calibration errors are shown in table~\ref{table:relative_error} in Appendix.}}
    \label{fig:migration_5Y_prediction_non_relocating}
\end{figure}

A naive approach suggests calculating the one-year migration matrix directly as 
\begin{equation*}
    p_{ij;\,1Y} = T_{ij;\,1Y}/x_i.
\end{equation*}
This, however, produces results which are inconsistent with the 5-year migration data. 
Indeed, the 5-year migration matrix is, by definition, $p_{ij;\,1Y} = \left[P^5\right]_{ij}$, where $P$ is the matrix of migration rates $p_{ij;\,1Y}$ and $\left[P^5\right]_{ij}$ stands for element in row $i$ and column $j$ of matrix $P^5$. The 5-year migration flow extrapolated from the 1-year migration flow is $\hat{T}_{ij;\,5Y} = p_{ij;\,5Y}\,x_i(t)$. Comparing the 5-year population obtained from actual migration  flow $\sum_{j \neq i} T_{ij;\,5Y}(2016)$ with the 5-year population obtained from the predicted migration flow $\sum_{j \neq i}\hat{T}_{ij;\,5Y}(2016)$, as shown in \figr{fig:migration_5Y_prediction_non_relocating}, we observe a systematic disagreement: the predicted numbers of movers are consistently higher than the actual numbers. In particular, in all Greater Capital Areas the average share of people who do not change their place of residence within 1 year is about $0.87-0.92$. The analogous share within 5 years is about $0.7-0.78$, while the predicted one is approximately $0.9^5 \approx 0.6$, as  shown in table \ref{table:stayer_share}. 

{We note that there exist, in principle, several alternative ways to calibrate the single-group model. They, however, give the same result: the single-group model is not capable of explaining the 5-years migration patterns from the 1-year migration patterns. We refer to Appendix  \ref{appendix:alternative_calibration} for the details of these calibrations.}

\begin{table}[h!]
\centering
    \caption{Share of people who do not change their place of residence (actual vs predicted). The values are based on 2016 Census data.}
    \vspace{3mm}
    \label{table:stayer_share}
    \begin{tabular}{|| c || c | c | c | c || } 
        \hline
        \multirow{2}{*}{GCA} & \multirow{2}{*}{1Y actual} & \multirow{2}{*}{5Y actual}  & 5Y predicted  & 5Y predicted \\
                             &                            &                             & 1-component   & 2-component \\
        \hline\hline
        Sydney & 0.909 & 0.733 & 0.630 & 0.735 \\
		\hline
		Melbourne & 0.905 & 0.734 & 0.618 & 0.736 \\
		\hline
		Brisbane & 0.889 & 0.702 & 0.569 & 0.704 \\
		\hline
		Adelaide & 0.911 & 0.752 & 0.635 & 0.754 \\
		\hline
		Perth & 0.895 & 0.71 & 0.584 & 0.712 \\
		\hline
		Hobart & 0.923 & 0.783 & 0.678 & 0.788 \\
		\hline
		Darwin & 0.874 & 0.713 & 0.527 & 0.717\\
		\hline
		Canberra & 0.904 & 0.720 & 0.615 & 0.722\\
        \hline
    \end{tabular}
\end{table}

In order to resolve this problem, we extend the model, assuming that the population comprises two groups instead of one, while staying with the general framework \eqref{eq:conductivty_components}. Each group is characterized by  its own relocation frequency, $\epsilon_1$ and $\epsilon_2$, which, in general, differ from each other. Furthermore, we restrict ourselves to the case where the population share of each group, $\alpha_1 \equiv \alpha$ and $\alpha_2 = 1 - \alpha$, is the same across all suburbs in the short-term and is equal to the total population share. If $\alpha$ is different for each suburb, the model will have an excessive number of parameters, which may improve the goodness of fit but will reduce the calibration robustness.
{We point out that the number of population groups with distinct relocation frequencies does not have to be equal to two: it simply has to differ from one. This is a crucial departure from a homogeneous population model which is not capable of explaining the actual relocation dynamics.
}

Within this framework, we deduce three parameters $\epsilon_1$ and $\epsilon_2$ from the data sets described above. There exist multiple estimation algorithms for similar models with a parametric structure of the matrices (see, e.g. \cite{Goodman1961, Cook2002, Frydman2018} for more details). Here we use a simple calibration technique by selecting parameters $\epsilon_1$, $\epsilon_2$ and $\alpha$ without specifying a parametric functional form for the elements of relocation matrix $H$.

We calculate migration flows $T_{ij}$ as the sum of two components:
\begin{equation}\label{eq:migration_flow_combined}
    \hat{T}_{ij;\,5Y}(2016) = x_i \left(\alpha  \left[P_1^5\right]_{ij} + (1-\alpha) \left[P_2^5\right]_{ij}\right),
\end{equation}
where $P_k=(1-\epsilon_k)I + \epsilon_k H$, $\left[P_k^5\right]_{ij}$ stands for element in row $i$ and column $j$ of the matrix $P^5_k$. Matrix $H$ is estimated as follows:
\begin{equation}\label{eq:H_estimate}
    h_{ij}=\frac{T_{ij;\,1Y}(2016)}{\sum_{k: k \neq i}T_{ik;\,1Y}(2016)},
\end{equation}
where $T_{ij;\,1Y}(2016)$ is the number of people migrated from suburb $i$ to suburb $j$ within one year period of 2015-2016. 
Relaxation rates $\epsilon_k$ and $\alpha$ can be found from the conditions \begin{equation}\label{eq:mobility_calibration}
\begin{array}{lclcl}
    \alpha (1-\epsilon_1) &+& (1-\alpha)(1-\epsilon_2) &=&s_{1Y}, \\
    \alpha (1-\epsilon_1)^5 &+& (1-\alpha)(1-\epsilon_2)^5 &=&s_{5Y},\\
    0 \leq \epsilon_1 \leq 1, &\quad& 0 \leq \epsilon_2 \leq 1,
\end{array}
\end{equation}
where $s_{1Y}$ is average share of stayers within one-year, $s_{5Y}$ is average share of stayers within  five-years period, which are calculated from the Census data. The actual values of $s_{1Y}$ and $s_{5Y}$ for the Australian Capital Areas are such that the solution to \eqr{eq:mobility_calibration} exists and unique. {It should also be noted that available data do not allow us to calibrate the value of $\alpha$, and thus it has to be fixed beforehand.} The general question of existence and uniqueness of the solution to \eqr{eq:mobility_calibration} with respect to $\alpha$ and $\epsilon_k$ is discussed in \secr{sec:two_components_calibration_analysis} in more details. 

The magnitudes of the migration outflow for each suburb predicted by this model for $\alpha=0.9$ are plotted against their actual magnitudes in  \figr{fig:migration_5Y_prediction_non_relocating} (numbers of movers, $\sum_{j \neq i}T_{ij;\,5Y}(2016)$). It is evident that the values predicted by the two-component model are in a stronger agreement with the actual data than the predictions obtained by the one-component model. 

The same analysis can be performed with the 2011 migration data and the corresponding predictions are shown in \figr{fig:migration_5Y_prediction_non_relocating_2011} in appendix.
Here, we again observe that the naive model produces a systematic bias in its predictions, while the two-component model provides a good fit to the data. From this comparison, we can conclude that the described methodology predicts the migration flows with a high precision, once the systematic bias produced by the one-component model is eliminated.

\subsection{{Intra-urban migration as diffusion}}\label{sec:diffusion}

{In this section, we describe intra-urban migration as an irreversible process of diffusion. We first follow the general description in \secr{sec:markov}, building the analogy for a general multi-component fluid. Next, we illustrate the analogy for a specific case of intra-urban migration in Sydney, using the results from \secr{sec:one_vs_two_components_data}.} 

The difference $U(t) \equiv X(t)-X_{eq}$ between the actual population at time $t$ and the equilibrium population $X_{eq}$ shows how far the system is away from equilibrium. Introducing the rate of population change $Q(t) \equiv X(t+1) - X(t)$ as the difference between two subsequent time steps and using \eqr{eq:stationary_total}, we can rewrite the population evolution equation \eqref{eq:forceflux1} as 
\begin{equation}\label{eq:forceflux}
    Q(t) = U(t) \cdot L
\end{equation}
where $L \equiv P-I$ and $I$ is the identity matrix. We view \eqr{eq:forceflux} as the central expression underlying {the analogy between intra-urban migration and diffusion. Indeed, if we consider two reservoirs with different fluid concentrations connected by a thin channel, there will exist a flow of fluid through that channel until these concentrations equilibrate. The rate of the concentration change in each of the reservoirs is proportional to the difference between the current concentration and the equilibrium concentration, with the proportionality coefficient being related to the diffusion coefficient, and following the same dependency as \eqr{eq:forceflux}. In general, \eqr{eq:forceflux} } has the form of a typical transport equation in non-equilibrium thermodynamics \cite{deGrootMazur}, which describes irreversible evolution of a thermodynamic system. For a closed system, this corresponds to a relaxation phenomenon, with $U(t)$ being the driving force, which drives the system towards equilibrium,  $Q(t)$ being the {rate of material change}. {Furthermore, $L$ is the matrix of transport coefficients, or simply the transport matrix, comprising the transport coefficients between each pair of suburbs}. The transport matrix determines how fast the system relaxes towards equilibrium. In context of urban dynamics, the irreversibility is ensured by constancy of the relocation matrix $P$: the fractions of residents migrating between two suburbs, $p_{ij}$, remain constant during relaxation (while the flows $T_{ij}$ and populations $x_i$ keep changing). In other words, once the equilibrium is reached, there is no driving force to reverse the relocation dynamics \eqref{eq:forceflux1}.

The transport coefficients are central in describing the irreversible evolution of a thermodynamic system. Similarly, the knowledge of the transport matrix is central in predicting the relocation dynamics in an urban system. An essential property of a transport coefficient in a physical system is that, in a closed system, it does not depend explicitly on time. This reflects the microscopic reversibility of molecular motion. While such principle does not exist \textit{a priori} for an urban system, {demanding that the transport matrix $L$ (and, therefore, the relocation matrix $P \equiv L + I$) does not change with time, indicates the microscopic reversibility of intra-urban relocation.} We will see below that this assumption is supported by actual data, helping us to derive the transport matrix from the Census data on relocation.

Substituting decomposition \eqref{eq:migration_matrix_decomposition} we obtain 
\begin{equation}\label{eq:conductivty}
    L = -\epsilon\,(I - H),
\end{equation}
so that the transport matrix factorizes into two terms. The temporal term, the coefficient $\epsilon$, shows the speed of relaxation towards equilibrium and characterizes the rate of the system irreversibility. The spatial term, the matrix $H$, shows the spatial distribution of a migration potential.

{We next point out that the population groups, introduced above, correspond to distinct components in a multi-component fluid mixture. Indeed, extending \eqr{eq:conductivty} to multiple population groups, we can write }
\begin{equation}\label{eq:conductivty_components}
    L_{k} = -\epsilon_{k}\,(I - H).
\end{equation}
{Here, the temporal term $\epsilon_{k}$ is different for each population group, corresponding to a fluid component with a distinct relaxation rate. In contrast, the spatial term $H$ is the same for all components,  corresponding to an external potential field.}
Writing the transport equation for each component separately, we obtain:
\begin{equation}\label{eq:forceflux_components}
    Q_{k}(t) = U_{k}(t)\cdot L_{k}
\end{equation}
where $L_{k}$ is the component-specific transport matrix defined by \eqr{eq:conductivty_components}, while $Q_k(t) \equiv X_k(t+1) - X_k(t)$ and $U_k(t) \equiv X_{k}(t)-X_{k,\,eq}$. 


With this analogy, the overall dynamics of intra-urban evolution follows a profile of diffusive relaxation. Specifically, at large $t$, the asymptotic decay of the driving force should be exponential, {with exponent $\lambda_k < 1$ being proportional to the second largest eigenvalue of matrix $H$. }
%
%
This implies that near equilibrium (when the values of $U_k(t)$ are small), the rate $Q_k(t)$ is asymptotically proportional to the driving force:
\begin{equation}\label{eq:equilibrium_2}
    \Vert Q_k(t) \Vert \sim (1-\lambda_k) \Vert U_k(t) \Vert.
\end{equation}
%

\begin{figure}[h!]
    \centering
    \includegraphics[height=3in]{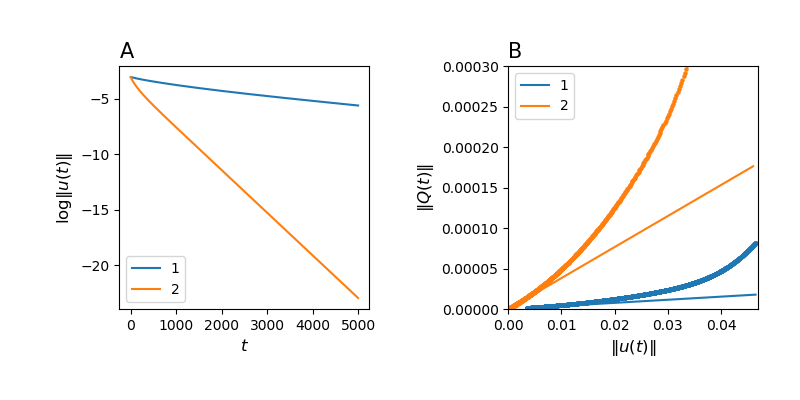}
    \caption{Exponential convergence of $U_k(t)$ {for each of the population groups, $k=1,2$}: (A). $\log \Vert U_k(t) \Vert$ is plotted against time step $t$; (B). $\Vert Q_k(t)\Vert$ is plotted against $\Vert U_k(t)\Vert$ (thick dotted curves) and the tangential lines with the slope $1-\lambda_k$ (solid straight lines){, where $\lambda_k$ is the second eigenvalue of the group relocation matrix}. For illustration purpose, both $U_k(t)$ and $Q_k(t)$ are normalized by total number of residents, $\alpha_k \overline{x}$, in the corresponding group.}
    \label{fig:diffusion_convergence}
\end{figure}

We illustrate the long-term relaxation dynamics for specific case of intra-urban migration in Sydney. As revealed in the previous subsection, there exist two population groups in Sydney, which correspond to two components in a fluid mixture.  \figr{fig:diffusion_convergence} shows the relaxation profiles for these components. Particularly, the left panel shows that that the driving force decays exponentially with time, as expected. Furthermore, the right panel shows that the near-equilibrium rate of relaxation is linearly proportional to the driving force, according to \eqr{eq:equilibrium_2} . This is, indeed, in agreement with the framework of linear irreversible thermodynamics \cite{deGrootMazur}, where the near-equilibrium {relaxation rate} is linearly proportional to the driving force, and the coefficient of proportionality is characterized by the second eigenvalue of the relocation matrix.

\subsection{{Predicting equilibrium population distribution}}\label{sec:long_term_structure_predictions}

\begin{figure}[h!]
\centering
    \includegraphics[height=7in]{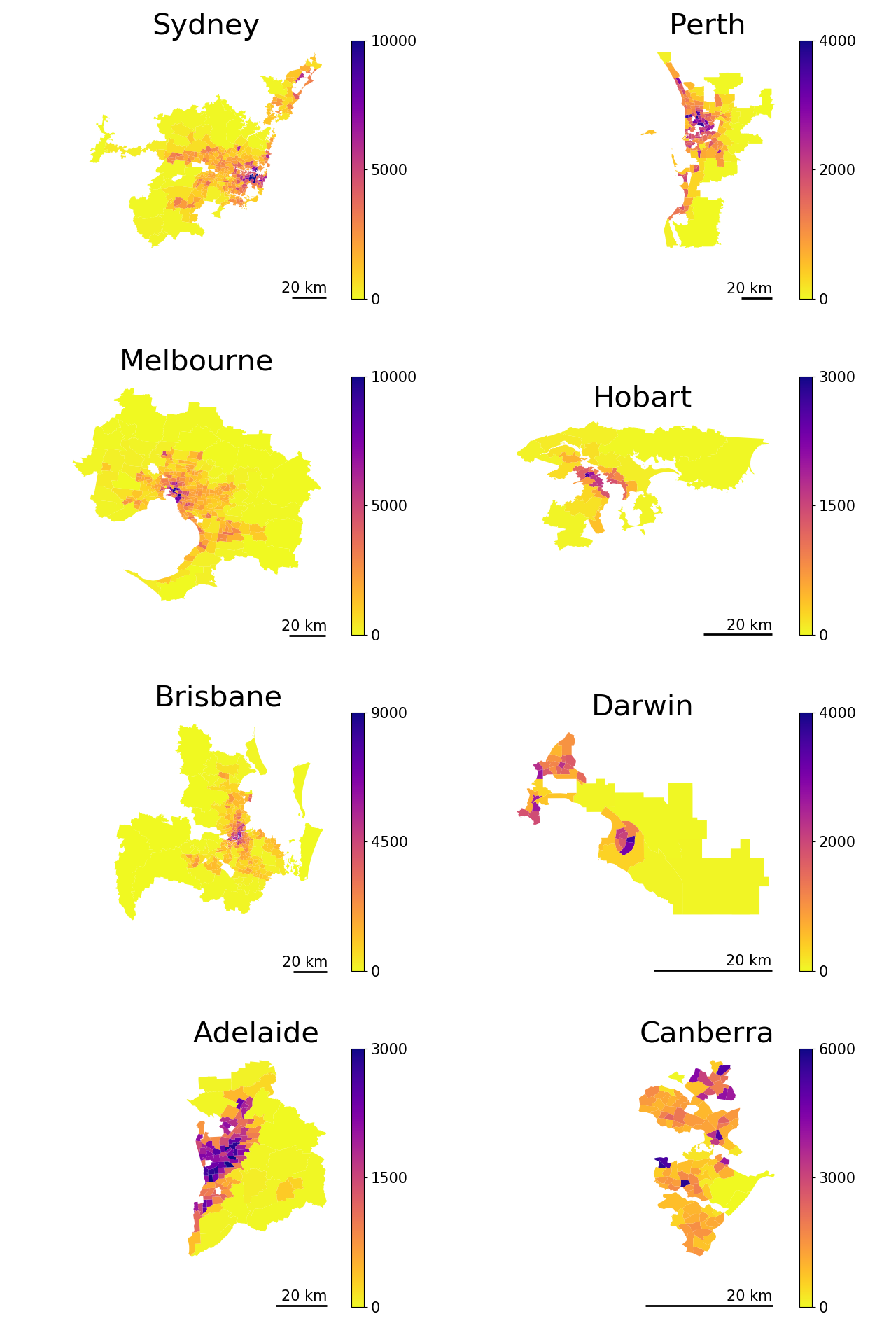}
    \caption{Long-run population structure prediction based on eigenvectors of migration matrix{, obtained from 2016 Census data. Scale bars in lower-left corners indicate distances equivalent to 20 km.}}
    \label{fig:population_prediction}
\end{figure}

{We next} build a long-term forecast for spatial structure of the Australian cities. {In doing so, we assume}  that the current migration flows  remain stable in the following years. 
As it has been mentioned above, the equilibrium structure $X_{eq}$ is independent of $\alpha$, $\epsilon_1$ and $\epsilon_2$; we calculate it as the first eigenvector of matrix $H$, {which} is obtained using \eqr{eq:H_estimate}.
The corresponding predictions are shown in \figr{fig:population_prediction}. To test the consistency of our predictions, we compare the predictions derived from 2016 data with the analogous predictions based on the 2011 data. \figr{fig:long_term_prediction_comparison} demonstrates that the outcomes based on 2011 and 2016 configurations are in a good agreement with each other. {This indicates that the relocation trends are stable in time, supporting the assumption about constancy of the relocation matrix $P$.} 

These results reveal that the equilibrium states of three out of eight capital cities (Sydney, Melbourne and Perth) are more spread out, compared with their current structure (shown in \figr{fig:actual_population} in appendix). The equilibrium structures of Brisbane, Adelaide and Darwin are similar to the current ones, while the structures of Hobart and Canberra are more compact than the current one.
These qualitative observations can be quantified by the spreading index method \cite{Louail2014,Volpati2018,Slavko2020}, which determines the degree of polycentricity and dispersal, as opposed to monocentricty and compactness of the city. The values of spreading index (calculated for both actual configurations of the Australian capital cities and for the predicted ones) are shown in table \ref{table:spreading_index}. {It is remarkable that our long-term prediction is independent of $\alpha$, $\epsilon$, and even the number of heterogeneous components (which can be larger than two in reality) as long as the spatial migration pattern described by matrix $H$ is shared by the entire population.}

\begin{table}[h!]
\centering
    \caption{Spreading index calculated for both current and predicted long-run structure of the Australian cities.}
    \vspace{3mm}
    \label{table:spreading_index}
    \begin{tabular}{|| c || c | c || } 
        \hline
        GCA 
        & Current 
        & Predicted \\
        \hline\hline
        Sydney & 0.29 & 0.74 \\ 
		\hline
		Melbourne & 0.26 & 0.43 \\ 
		\hline
		Brisbane & 0.32 & 0.37 \\ 
		\hline
		Adelaide & 0.54 & 0.53 \\ 
		\hline
		Perth & 0.44 & 0.58 \\ 
		\hline
		Hobart & 0.38 & 0.25 \\ 
		\hline
		Darwin & 0.75 & 0.79 \\ 
		\hline
		Canberra & 1.06 & 0.92 \\
        \hline
    \end{tabular}
\end{table}

\section{Robustness evaluation}\label{sec:robustness}
In the previous section, we have shown that our non-equilibrium framework of diffusive intra-urban relaxation explains the short-term migration data and is able to provide long-term predictions. The important element of the model is the assumption that the population comprises multiple components with different relocation frequencies, which, in the context of our framework, correspond to different relaxation rates. In this section, we investigate {robustness of this claim}, analysing the extent of its applicability. In particular, in \secr{sec:two_components_calibration_analysis} we study the two-component model, arguing that this case is sufficient to consistently describe the short-term migration. In \secr{sec:heterogeneous_H} and \secr{sec:heterogeneous_sydney}, we explore the sensitivity of the equilibrium configuration to the spatial migration patterns (captured by matrix $H$), varying between different population components. In \secr{sec:heterogeneous_H} we do this for an abstract city with extreme migration patterns, while in \secr{sec:heterogeneous_sydney} we extend this analysis to a specific case (Sydney).

%
\subsection{Heterogeneity of the population}
{In \secr{sec:one_vs_two_components_data}, we reported that the baseline one-component model systematically predicts higher rates of five-year migration than observed in the reality. Here we show that having a homogeneous population mobility can not produce consistent migration predictions; hence, the population has to be heterogeneous with respect to its mobility. The population, which is comprised of two groups with two distinct relocation frequencies is a minimal realization of such heterogeneity.
}

\begin{figure}[h!]
\centering
    \includegraphics[height=2.7in]{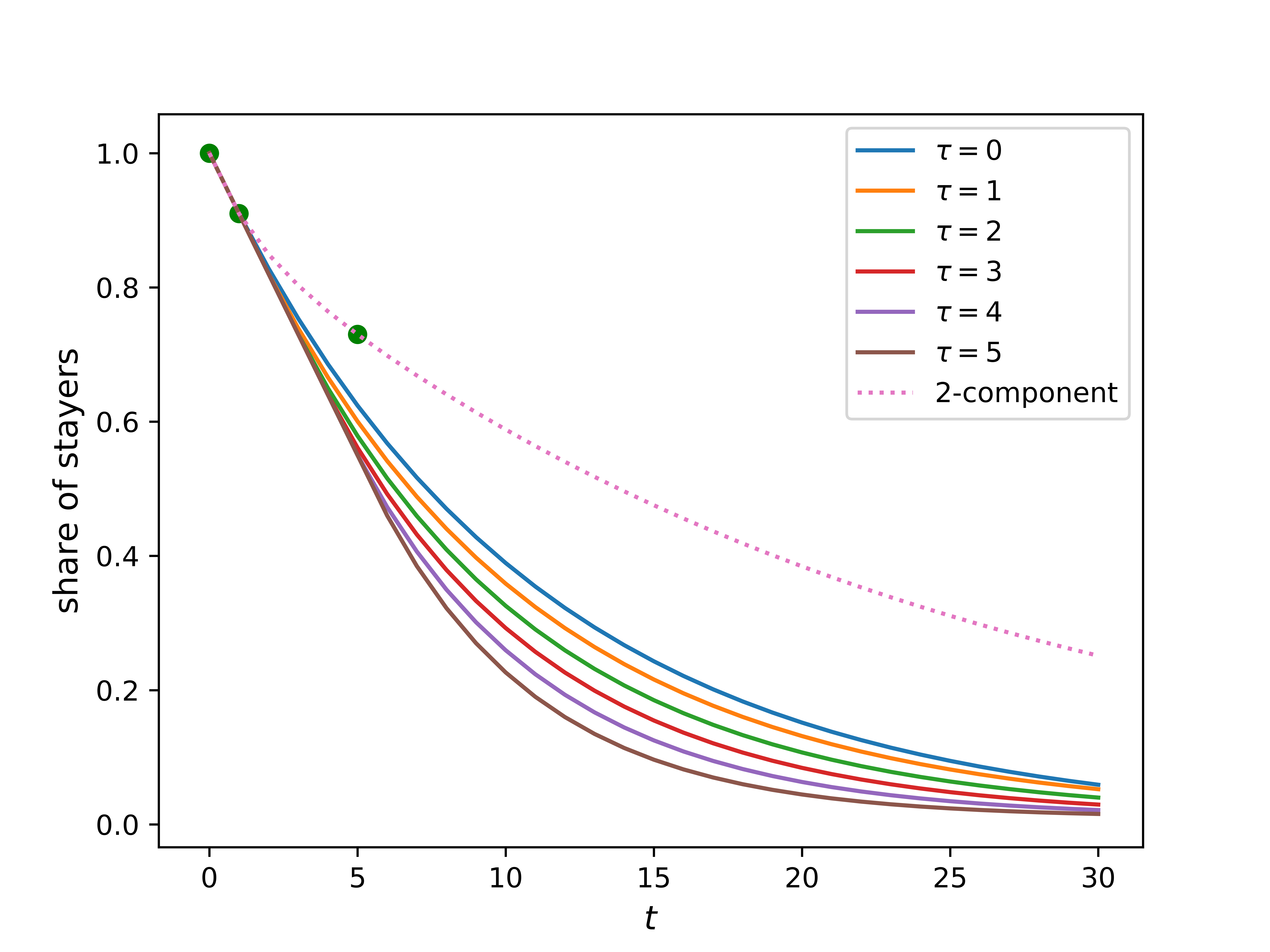}
    \caption{The share of people who do not change their place of residence within period $t$ plotted against the length of this period. Green dots correspond to the actual values for Sydney ($s_{0Y}=1$; $s_{1Y}=0.91$; $s_{5Y}=0.73$). Solid curves correspond to the model where people do not relocate within $\tau$ years after their last relocation ($\tau$ ranges from 0 to 5), see \eqr{eq:migration_5Y_memory}. The dotted curve corresponds to the two component model ($\alpha = 0.9$), see \eqr{eq:migration_flow_combined}. All models are calibrated to the Sydney relocation data. All solid curves pass through the actual one-year relocation rate $s_{1Y}=0.91$ but go well below the corresponding five-year value, $s_{5Y}=0.73$.}
    \label{fig:stayer_curve_with_memory}
\end{figure}

{One may expect that the 5-years relocation rate, observed in reality, is lower than the one predicted from the 1-year relocation rate, due to the low mobility of recently relocated people. Indeed, if an individual has moved into a new home recently, there might be not much incentive for them to move further relatively quickly. To represent this constraint, we assume that people do not relocate for $\tau$ years after their last relocation (immobility assumption), and calculate the share of those who have not relocated for different values of this parameter, $\tau$. \figr{fig:stayer_curve_with_memory} compares how this share changes in time for homogeneous populations (full solution of this model is provided in \secr{appendix:memory_model} in Appendix) and a two-component population without immobility assumption. It is evident from the figure, that the immobility assumption does not improve the model. Indeed, no value of the immobility period $\tau$ can match the five-year relocation flow. In such setting, presence of the residents that do not relocate within a certain period, implies a higher relocation frequency for those who do. This results in the 5-year relocation rate predicted by the homogeneous model with the immobility assumption to be higher than that produced by the simple homogeneous model (the baseline model), leading to an outcome opposite to the one that was expected.}

{This analysis shows that a lower mobility of the recently relocated people cannot be a valid explanation for the lower actual 5-year relocation rate. Therefore, we  conclude that having a homogeneous population comprised of only one component is not sufficient to make the model consistent with the data.}

\subsection{Solution space of the two-component model}\label{sec:two_components_calibration_analysis}

In \secr{sec:one_vs_two_components_data}, we have calculated $\epsilon_1$ and $\epsilon_2$ using the average shares of stayers within one-year ($s_{1Y}$),  five-years period ($s_{5Y}$) and conditions \eqref{eq:mobility_calibration}. 
The values of $\epsilon_1$ and $\epsilon_2$ depend on $\alpha$, which is not known without specifying the nature of the groups. This, however, does not affect the possibility to split the population in two groups and  obtain consistent predictions of the five-year migration patters from the one-year migration patterns.

\begin{figure}
\centering
    \includegraphics[width=\textwidth]{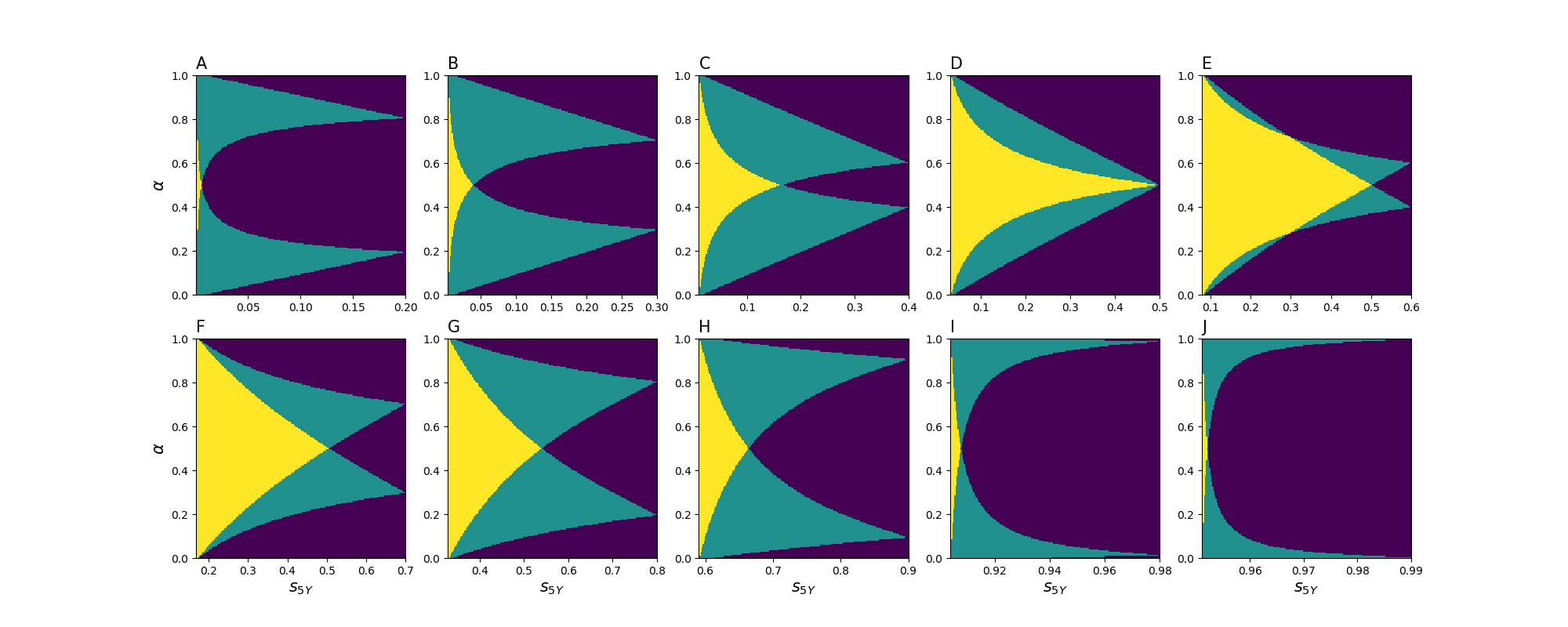}
    \caption{Number of solutions to \eqref{eq:mobility_calibration} depending on $\alpha$ and $s_{5Y}$ for A. $s_{1Y}=0.2$, B. $s_{1Y}=0.3$, C. $s_{1Y}=0.4$, D. $s_{1Y}=0.5$, E. $s_{1Y}=0.6$, F. $s_{1Y}=0.7$, G. $s_{1Y}=0.8$, H. $s_{1Y}=0.9$, I. $s_{1Y}=0.98$, J. $s_{1Y}=0.99$. Yellow areas correspond to two distinct solutions, green areas represent one solution, dark purple stands for no solution.}
    \label{fig:calibration_diagram}
\end{figure}

The non-linear system of  algebraic equations \eqref{eq:mobility_calibration} allows one to calculate the relocation frequencies $\epsilon_1$, $\epsilon_2$ for a given composition $\alpha$. It consists of two equations while containing 3 unknown variables ($\epsilon_1$, $\epsilon_2$ and $\alpha$),
and for a given $\alpha$ it can have up to 5 real roots. Some of the roots may not belong to the range from 0 to 1 and, therefore, cannot be valid solutions for $\epsilon_1$ and $\epsilon_2$. \figr{fig:calibration_diagram} demonstrates that, depending on $s_{1Y}$, $s_{5Y}$ and $\alpha$, there exist 2, 1 or 0 solutions. Furthermore, it is evident from \figr{fig:calibration_diagram}, that for any $s_{5Y}$ ranging from $s_{1Y}^5$ to $s_{1Y}$ there exists at least one solution for the pair $\epsilon_1$ and $\epsilon_2$. This means that it is always possible to calibrate the model \eqref{eq:mobility_calibration} as long as $s_{5Y} \geq s_{1Y}^5$ (the other inequality, $s_{5Y} \leq s_{1Y}$ holds automatically), which is the case for the actual migration data. 

\begin{figure}
\centering
    \includegraphics[height=2.7in]{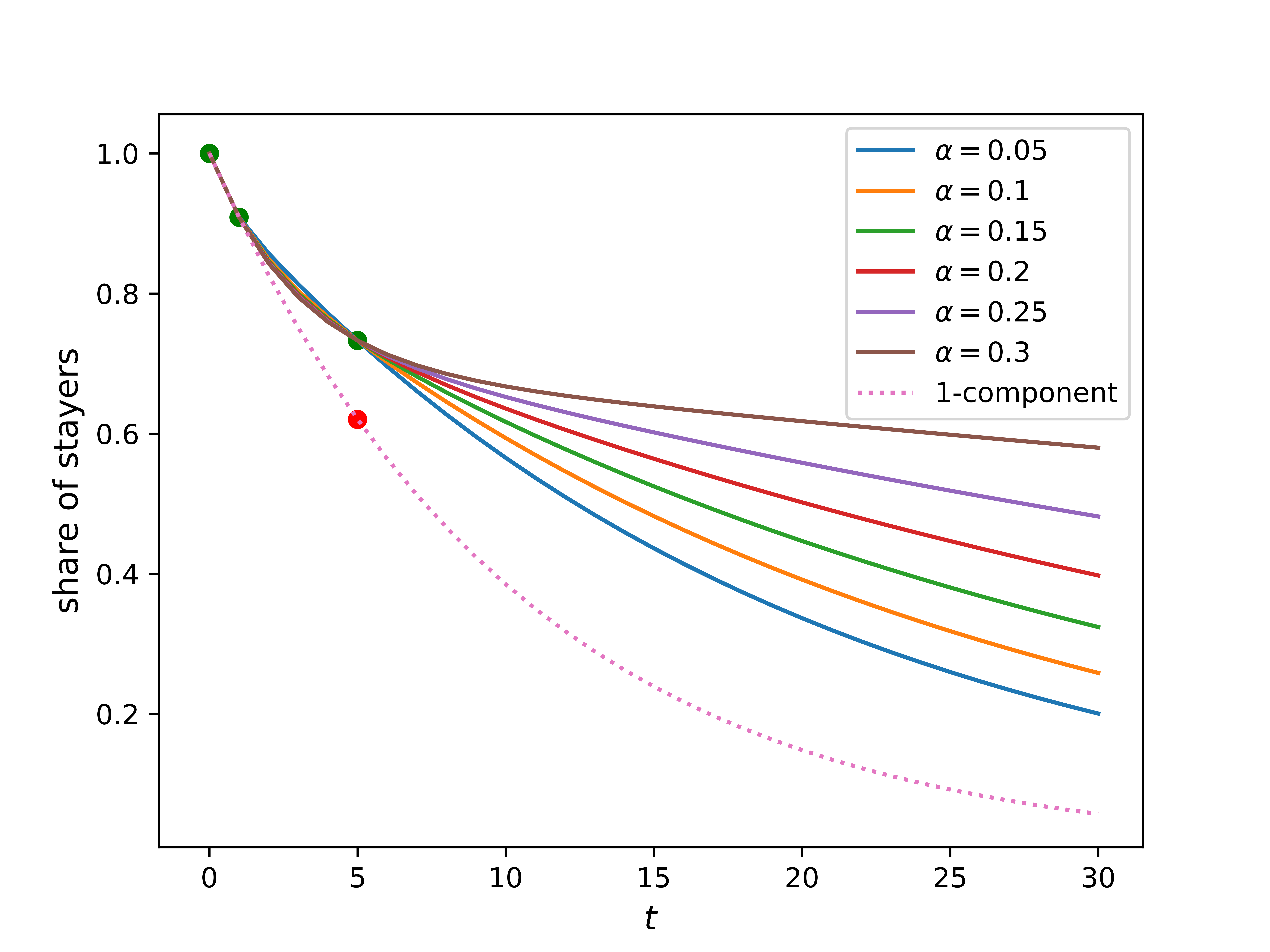}
    \caption{The share of people who do not change their place of residence within period $t$ plotted against the length of this period. Dotted curve corresponds to the naive single-component model (calibrated to one-year value, $s_{1Y}=0.91$). Solid lines describe a family of the two-component model predictions matching actual one-year and five-year values (Sydney values, $s_{1Y}=0.91$ and $s_{5Y}=0.73$, are taken as an example) for different levels of $\alpha$. All solid curves pass through 3 common points (green): $s_{0Y}=1$; $s_{1Y}=0.91$; $s_{5Y}=0.73$. The dotted curve passes through the first two green points and its five-year prediction is marked in red.}
    \label{fig:forward_stayer_curve}
\end{figure}

Although it is not feasible to estimate $\alpha$ directly from the current data set, it would be possible to do so with a longer record of internal migration. In particular, if we also knew people's places of residence 10 years ago, 15 years ago etc., we would be able to determine the value of $\alpha$, which predicts the share of movers and stayers with a higher precision. {This idea is illustrated in \figr{fig:forward_stayer_curve}, which demonstrates the stayer share values predicted by the two-component model. Parameters $\epsilon_1$ and $\epsilon_2$ are calibrated to $s_{1Y}=0.91$ and $s_{5Y}=0.73$ (Sydney values are used as an example). Values of $\alpha$ vary from 0.7 to 0.95 (solution of \eqref{eq:mobility_calibration} exists and is unique if $0.05 \leq \alpha \leq 0.33$ and $0.67 \leq  \alpha \leq 0.95$ but the solutions for $\alpha = a$ and $\alpha = 1 - a$ are equivalent due to symmetry).}
For the 30 years horizon, the predictions for the share vary from 0.2 (if $\alpha=0.95$) to 0.6 (if $\alpha=0.7$). This means that the two-component model can consistently calibrate a larger variety of data than the one-component model. 
If, however, the actual structure of the population is more complex, e.g., is made of a larger number of components, the two-component model would not be able to adequately account for the corresponding data.

\subsection{Component-specific relocation matrix}\label{sec:heterogeneous_H}

As has been shown previously, heterogeneity in mobility rates $\epsilon_k$ does not affect the equilibrium structure of the city as long as all groups have the same relocation matrix $H$. This assumption, however, may not be valid if we do not observe $H$ for each group directly. This might be important, as there exists an empirical evidence that different social groups (such as renters and mortgagors) have different migration patterns \cite{Crosato2020}. Thus, it is important to assess the possibility for the matrices $H_k$ to be component-specific, or in other words, heterogeneous.

If matrices $H_k$ are heterogeneous, the equilibrium population structure $X_{eq}$ is no longer independent of the compositions $\alpha_k$ and relocation rates $\epsilon_k$. In particular, matrices $H_k$ may have different stationary vectors $X_{k,eq}$, in which case it is impossible to estimate the stationary population structure unless matrices $H_k$ are observed directly. The equilibrium structure $X_{eq}=\sum_{k=1}^{C} X_{k,eq}$ depends on the individual matrices $H_k$ and cannot be expressed via the aggregated relocation matrix $\hat{H} = \sum_{k=1}^{C} \alpha_k H_k$.

In this section, we show that the structure calculated using the aggregated matrix $\hat{H}$ can still give a reasonable approximation for the stationary population structure $X_{eq}$, even if the individual vectors $X_{k,eq}$ differ drastically. 
To demonstrate this, we consider two artificial examples. In the first example, the population components $1$ and $2$ generate the migration flows with opposite directions. In the second example, there are two groups of suburbs (A and B), and the members of component $1$  always relocate to the suburbs within $A$, while the members of component $2$  always relocate to suburbs within $B$. These two examples are considered for a linear toy city comprising 99 suburbs. All suburbs are located along a line, such that suburb 50 is the ``central'' one. 

\begin{figure}
\centering
    \includegraphics[height=7in]{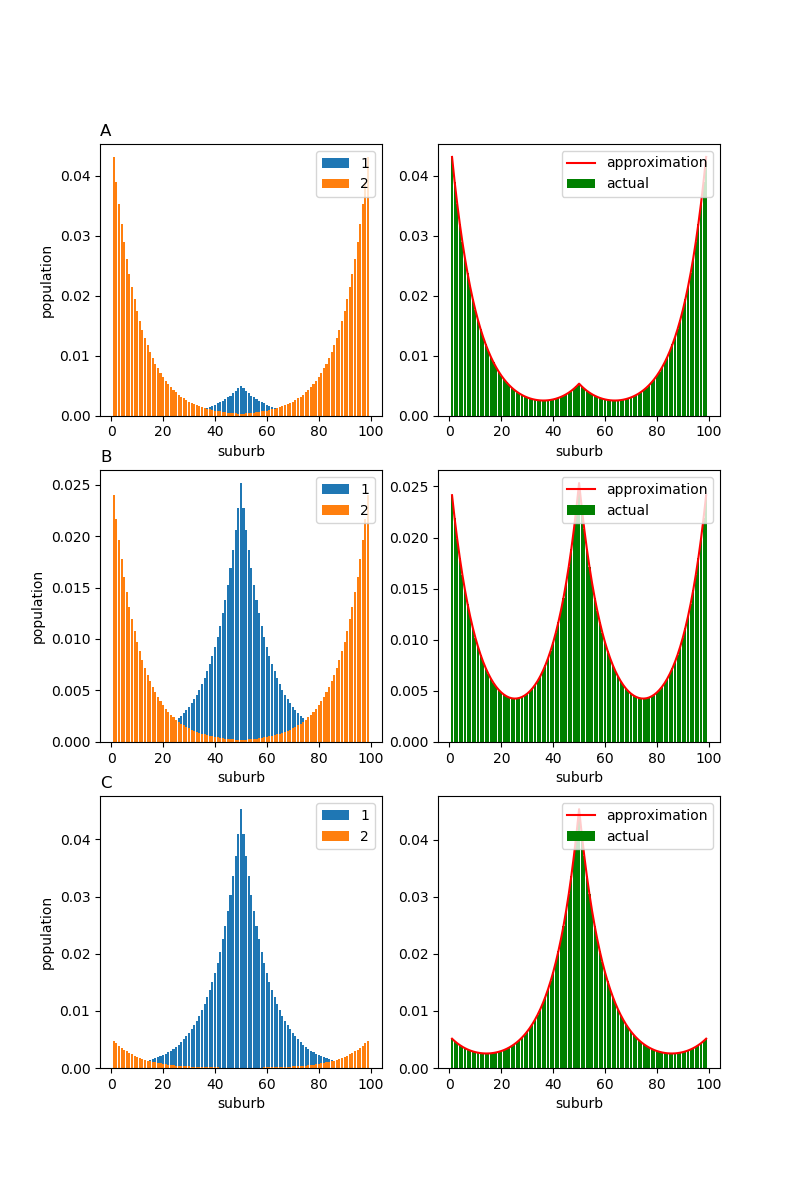}
    \caption{Stationary population structure for the case where components 1 and 2 have migration flows with opposite directions: (A) $\alpha = 0.1$; (B) $\alpha = 0.5$; (C) $\alpha = 0.9$. {Componentwise population structure is show in the left column. Total population structure is shown in the right column. For all values of $\alpha$, the approximations $\hat{X}_{eq}$ obtained from the \emph{observable} matrix $\Hat{H}$ (red solid line) are almost indistinguishable from the ground-truth equilibria $X_{eq}$ (green bars).}}
    \label{fig:heterogeneous_H_opposite_flows_example}
\end{figure}

In the first example, matrix $H_1$ consists of elements $h_{1;\,ij}$ given by:
\begin{equation}
    h_{1;\,ij} = \frac{e^{-\beta(d_j-d_i)}}{\sum_{k=1}^{99} e^{-\beta(d_k-d_i)}},
\end{equation}
where $d_i=|i-50|$ is the distance from $i$ to the ``central" suburb (suburb $50$), $\beta=0.1$. Elements of $H_2$ are defined as follows:
\begin{equation}
    h_{2;\,ij} = \frac{e^{-\beta(d_i-d_j)}}{\sum_{k=1}^{99} e^{-\beta(d_k-d_j)}}.
\end{equation}
This form of $H_1$ and $H_2$ means that the members of component 1 prefer to relocate to more central suburbs (that are close to the suburb 50), while the members of component 2 relocate to the peripheral suburbs (which are far from the suburb 50) more frequently. The equilibrium population distributions $X_{1,eq}$ and $X_{2,eq}$ are displayed in \figr{fig:heterogeneous_H_opposite_flows_example} (left column). As one might have anticipated, the population of component $1$ forms a monocentric structure around the ``central'' suburb, while the population of component $2$ predominantly inhabits the peripheral suburbs. The corresponding total population structure $X_{eq}$ and its approximation $\hat{X}_{eq}$ obtained from matrix $\hat{H} = \sum_{k=1}^{C} \alpha_k H_k$ is shown in the right column. In all three cases: (A) $\alpha = 0.1$; (B) $\alpha = 0.5$; (C) $\alpha = 0.9$, the actual population structures $X_{eq}$ {(green bars)} lie very close to the corresponding approximations $\hat{X}_{eq}$ {(red solid line)}.

In the second example, we fill columns 26-74 of matrix $H_1$ with positive random numbers and the other columns are filled with zeros. In contrast, to fill matrix $H_2$, we assign zero values to columns 26-74 and positive random numbers to columns 1-25 and 75-99. Each row in both matrices is normalized so that its elements sum to one.

\begin{figure}
\centering
    \includegraphics[height=7in]{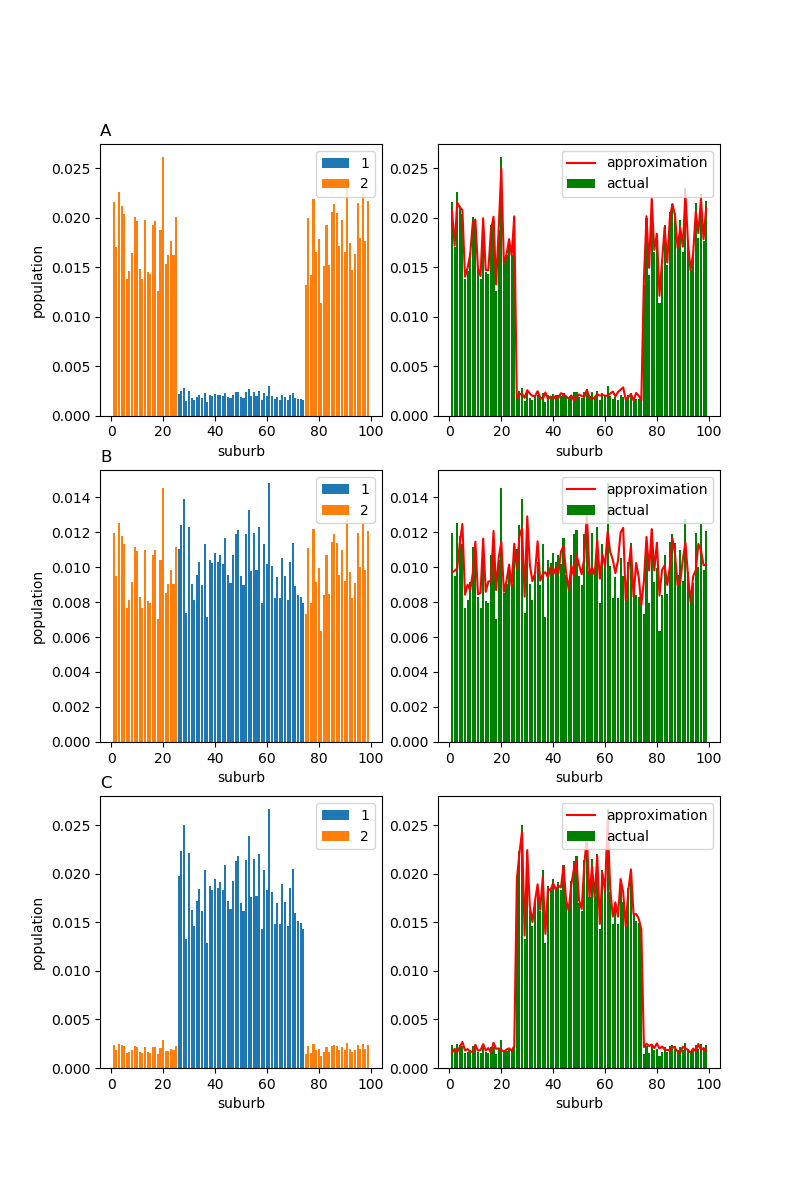}
    \caption{Stationary population structure for the case where the first group members always relocate to the central districts while the second group members migrate to the peripheral ones: (A) $\alpha = 0.1$; (B) $\alpha = 0.5$; (C) $\alpha = 0.9$. {Componentwise population structure is show in the left column. Total population structure is shown in the right column. For all values of $\alpha$, the approximations $\hat{X}_{eq}$ obtained from the \emph{observable} matrix $\Hat{H}$ (red solid line) are almost indistinguishable from the ground-truth equilibria $X_{eq}$ (green bars).}}
    \label{fig:heterogeneous_H_segregated_flows_example}
\end{figure}

It is natural to anticipate that, in the equilibrium, all members of component $2$ will live in suburbs 26-74 while the members of component $1$ will live in suburbs 1-25 and 75-99 (left column in \figr{fig:heterogeneous_H_segregated_flows_example}; cases A, B and C correspond to $\alpha=0.1;\quad 0.5;\quad 0.9$ respectively). {In the right column of \figr{fig:heterogeneous_H_segregated_flows_example}, we again observe that, regardless of $\alpha$, the actual equilibrium structure $X_{eq}$ (green bars) does not deviate significantly from its approximation $\hat{X}_{eq}$ (red solid line).}

\subsection{Sydney case study}\label{sec:heterogeneous_sydney}

To demonstrate the robustness of the results presented in \secr{sec:long_term_structure_predictions} with respect to the heterogeneity of relocation patterns, we extend this analysis to Greater Sydney  Capital Area. 
In a real city, the migration matrices $H_1$ and $H_2$ are not normally observed separately. Moreover, these matrices cannot be assigned arbitrarily, as they need to be consistent with the actual migration data. In particular, following the procedure suggested in \secr{sec:heterogeneous_H}, the component-specific matrices have to be defined such that $\alpha H_1 + (1-\alpha)H_2 = H$, with $H_1$ accounting for the relocations flowing primarily into central districts, and $H_2$ corresponding to the relocations flowing primarily into the peripheral areas.

To accomplish this task, we choose a distance threshold $\overline{d}$, and select suburb groups $A(\overline{d})$ and $B(\overline{d})$ so that $A(\overline{d})$ contains only suburbs with the distance to central business district being less than $\overline{d}$, while $B(\overline{d})$ contains the rest of the suburbs. Next, we assume that when relocating, the members of component $1$ almost always choose suburbs from set $A(\overline{d})$ while group $2$ members choose suburbs from set $B(\overline{d})$. 
Finally, we calibrate $H_1$ and $H_2$ to actual relocation data denoting $a_i \equiv \sum_{j:d_j \leq \overline{d}}\,h_{ij}$ in each row $i$ and define elements $h_{1;\,ij}$ of $H_1$ as follows:
\begin{equation*}
    h_{1;\,ij}=
    \begin{cases}
        \displaystyle \frac{h_{ij}}{\alpha}, \quad \text{if } d_j \leq \overline{d},\\\\
        \displaystyle \frac{\alpha-a_i}{\alpha\left(1-a_i\right)}h_{ij}, \quad \text{if } d_j > \overline{d},
    \end{cases}
\end{equation*} 
if $a_i \leq \alpha$, and
\begin{equation*}
    h_{1;\,ij}=
    \begin{cases}
        \displaystyle \frac{h_{ij}}{a_i}, \quad \text{if } d_j \leq \overline{d},\\\\
        0, \quad \text{if } d_j > \overline{d},
    \end{cases}
\end{equation*} 
if $a_i > \alpha$. The elements of $H_2$ are then given by: 
\begin{equation*}
    h_{2;\,ij}=\frac{1}{1-\alpha}(h_{ij}-\alpha \,h_{1;ij}).  
\end{equation*}
In other words, all members of the first component 
move to areas inside $A(\overline{d})$ and all members of the second component move to areas inside $B(\overline{d})$, but the total share $a_i$ of people from $i$ who move to suburbs inside $A(\overline{d})$ might differ from $\alpha$. If $a_i \leq \alpha$, we assume that all the people who relocate from $i$ to $A(\overline{d})$ belong to the first component and so does the proportion $(\alpha-a_i)/(1-a_i)$ of the people migrating to the other suburbs; while the rest of the $i$'s residents belong to the second component. Conversely, if $a_i > \alpha$, we assume that only the proportion $\alpha/a_i$ of those who relocate to $A(\overline{d})$ belong to the first component, while others belong to the second component. 

\begin{figure}
\centering
    \includegraphics[height=6in]{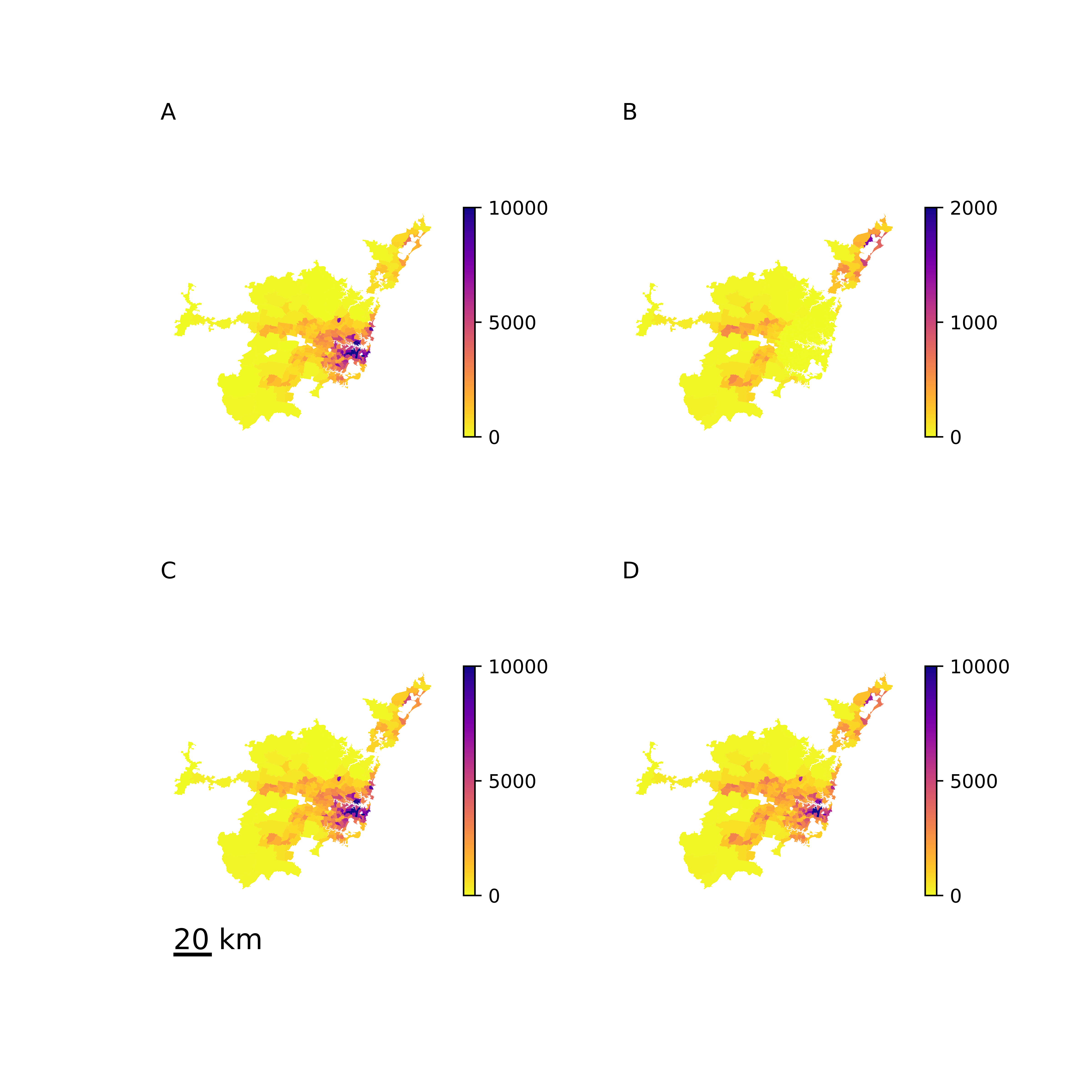}
    \caption{Equilibrium population density in Sydney for the case of heterogeneous relocation matrices $H_k$. (A) first group equilibrium structure $X_{1,eq}$; (B) second group equilibrium structure $X_{2,eq}$; (C) total equilibrium structure $X_{eq}$; (D) approximation $\hat{X}_{eq}$ obtained from the overall relocation matrix $H$. {The scale bar in lower-left corner indicates the distance equivalent to 20 km.}}
    \label{fig:Sydney_heterogeneous_H}
\end{figure}

It is easy to see that in that case $\alpha H_1 + (1-\alpha)H_2 = H$, and that all elements $h_{1;\,ij}$ and $h_{2;\,ij}$ are positive and in each row $i$, we have $\sum_{j=1}^N h_{1;\,ij}=1$ and $\sum_{j=1}^N h_{2;\,ij}=1$, which is recquired by construction.
The resulting equilibrium structure of the population density is shown in \figr{fig:Sydney_heterogeneous_H} for $\alpha=0.9$, $\overline{d}=22$ km (median distance to the central business district). Similarly to the previous examples, the approximated $\hat{X}_{eq}$ (\figr{fig:Sydney_heterogeneous_H}D) does not differ significantly from the actual value $X_{eq}$ (\figr{fig:Sydney_heterogeneous_H}C) although $X_{1,eq}$ and $X_{2,eq}$ do differ drastically (\figr{fig:Sydney_heterogeneous_H}A and B) .

From these examples, we can conclude that heterogeneity in matrices $H_k$ has a limited effect on the long-term population structure and it is possible to obtain an accurate prediction by using only the aggregate relocation matrix $H = \sum_{i=1}^{C} \alpha_k H_k$.





\section{Conclusions}\label{sec:conclusions}

We have introduced a diffusive migration framework, which describes intra-urban migration as an irreversible evolution of the urban population. The results have been tested for residential relocation data available from the Australian Census for eight Greater Capital areas over 10 years.

Using this framework we were able to {explain} the medium-term (5 years) migration patterns from the short-term (1 year) migration patterns. 
We have shown that this is possible to achieve only if the population is not homogeneous and has an internal structure. In particular, such population should be comprised of at least two components, with each component having a distinct relocation frequency. Such relocation frequency corresponds to a particular relaxation time of the component.

This heterogeneity of migration frequencies has an intuitive interpretation. For example, the group of residents which migrate more often can be interpreted as renters (who are less attached to their  place of residence, and are relatively free to change it as soon as they identify a better option) and home-owners (for whom it may be more problematic to change the place of residence due to the transaction costs and peculiarities of the housing market and individual circumstances).

Using this diffusive migration framework, we produced a long-term prediction for the Australian capital cities' structures, based on the short-term migration data, with the only assumption about the temporal stability of the migration rates. According to our predictions, the largest capital cities (Sydney, Melbourne and Perth) are moving towards more spread-out configurations, while Hobart and Canberra exhibit a more compact structure in the equilibrium.
The other capitals, Brisbane, Adelaide and Darwin, are likely to preserve their current configuration in the long-run. These results are consistent with the previous studies predicting the possibility of polycentric transition in Sydney and Melbourne \cite{Crosato2018,Crosato2020}.

Our predictions are robust with respect to the composition of the migration components, as well as  possible heterogeneity of their relocation patterns, both dynamic and spatial. In particular, we have  analytically shown that the long-run equilibrium is independent of the size of each community and their relocation rates. The robustness with respect to spatial heterogeneity of relocation has been shown numerically through an abstract illustrative example. In this example, the relocation communities have opposite preferences regarding their destination: members of the first group  prefer central districts, while their counterparts prefer the peripheral ones.

The temporal stability of the migration flows is a crucial element of our long-term analysis. Despite being consistent within the period of observation (2006--2016), they may be affected by multiple factors in future: the human migration is a complicated non-linear process involving multiple interdependent factors, often leading to various phase transitions and critical phenomena \cite{Weidlich1988,Harris1978,Wilson2008,Osawa2017,Crosato2018,Dynamic_resettlement_paper,Crosato2020}. However, the relocation data may contain some unique features that are not captured in other static human mobility and land use data (e.g., \cite{Clarke1998,Schneider2013,Crosato2018,Ellam2018,Crosato2020}). Thus, we believe that the proposed dynamic framework for the intra-urban migration, enabling robust long-term predictions, offers a principled approach to modeling out-of-equilibrium urban development.

\vspace{6pt} 




\textbf{Author Contributions.} Conceptualization, B.S., M.P. and K.G.; methodology, B.S., M.P. and K.G.; software, B.S.;  formal analysis, B.S. and K.G.; data curation, B.S.; writing--review and editing, B.S., M.P. and K.G.; supervision, K.G. and M.P. 

\textbf{Funding.} This work was funded by the Australian Research Council Discovery Project \\ DP170102927.


\textbf{Conflicts of Interest.} The authors declare no conflict of interest. The funders had no role in the design of the study; in the collection, analyses, or interpretation of data; in the writing of the manuscript, or in the decision to publish the results.

\bibliographystyle{elsarticle-num-names}
\bibliography{bibliography}




\newpage
\appendix

\section{Mathematical derivations}

\subsection{Convergence to the same equilibrium in (\ref{eq:conductivty_components})}\label{appendix:same_equilibrium}
\textbf{Proposition.} \textit{If in equation \eqref{eq:conductivty_components} each component $k$ has a unique equilibrium structure $X_{k,eq}$, then it is independent of $\epsilon_k$ and has the form $X_{k,eq} = \alpha_k X_{eq}$ where $X_{eq}$ can be found as a solution of left eigenvector of the matrix $H$ that corresponds to the unit eigenvalue.}

\textit{Proof.} By definition of equilibrium $X_{k,eq}$ we obtain:
\begin{equation}
    X_{k,eq} \left((1-\epsilon_k)I+\epsilon_k H\right) = X_{k,eq}, 
\end{equation}
which is equivalent to
\begin{equation}\label{eq:homogeneous_system}
    \epsilon_x X_{k,eq} \left(H-I\right) = \theta,
\end{equation}
where $\theta=(0, 0, \cdots, 0)$ is a zero row vector.

Homogeneous systems of equation \eqref{eq:homogeneous_system} are independent of the constant multipliers $\epsilon_k$ and therefore are identical and equivalent to
\begin{equation}\label{eq:main_homogeneous_system}
    v(H-I)=\theta.
\end{equation}
This implies that all $X_{k,eq}$ are eigenvectors of $H$ that correspond to the unit eigenvalue. Exact values of $X_{k,eq}$ can be found from the constraint on total population (each component $k$ has total population $\alpha_k \overline{x}$). If we choose a vector $v$ satisfying \eqref{eq:main_homogeneous_system} and whose elements sum to the total population $\overline{x}$ ($\sum_{i=1}^{N} v_i=\overline{x}$), it is easy to see that $X_{k,eq} = \alpha_k v$ for each component $k$ and the total equilibrium structure is given by $X_{eq} = \sum_{k=1}^{C}X_{k,eq} = v$.

\subsection{Derivation of equation (\ref{eq:mobility_calibration})}\label{appendix:calibration}
Equation \eqref{eq:mobility_calibration} is a consequence of \eqref{eq:migration_flow_combined} which can be rewritten as
\begin{equation}
    \begin{cases}
        P_{1Y} = \alpha P_1 + (1-\alpha)P_2,\\
        P_{5Y} = \alpha P_1^5 + (1-\alpha)P_2^5.
    \end{cases}
\end{equation}
where $P_{1Y}$ and $P_{5Y}$ are the aggregate one-year and five-year migration rate matrices respectively.
To obtain the first equation in \eqref{eq:mobility_calibration}, we directly apply decomposition \eqref{eq:migration_matrix_decomposition} and equate the diagonal elements (without loss of generality, we set $h_{ii}=0$). To obtain the second one, we take into account that elements $h_{ij}$ are small and have order of magnitude of $(1/N) \ll 1$. This implies that all small powers of $H$ have elements with the same order of magnitude, which can be disregarded. In particular, the elements of $H^2$ are $h_{ij;\,2Y}=\sum_{k=1}^N h_{ik;\,1Y} h_{kj;\,1Y} \sim 1/N$, the elements of $H^3$ are $h_{ij;\,3Y} = \sum_{k=1}^N h_{ik;\,1Y} h_{kj;\,2Y} \sim 1/N$ and, hence, the elements of $H^5$ are $h_{ij;\,5Y}=\sum_{k=1}^N h_{ik;\,1Y} h_{kj;\,4Y} \sim 1/N$ (symbol $\sim$ means ``has order of magnitude of''). Hence diagonal elements of both matrices $P_1^5$ and $P_2^5$ are approximately $\approx (1-\epsilon_k)^5$.

\subsection{{Migration model with memory}}\label{appendix:memory_model}
{We let $a_i$ denote the share of people relocated $i$ years ago ($i=1,2,\dots \tau$). We use $a_0$ to represent share of people who relocated more than $\tau$ years ago or did not relocate at all.
Stationary values of $a_0,a_1,\dots a_{\tau}$ can be found from the following equations:}
\begin{equation}
    \begin{cases}
        a_1 = \epsilon a_0,\\
        a_2 = a_1,\\
        \dots\\
        a_{\tau} = a_{\tau-1},\\
        a_0 + a_1 + \dots + a_{\tau} = 1.
    \end{cases}
\end{equation}
{The solution is:}
\begin{equation}
    a_0 = \frac{1}{1+n\epsilon}, \quad\quad a_1 = a_2 = \dots = a_{\tau} = \frac{\epsilon}{1+n\epsilon}
\end{equation}
{To match one-year relocation rate, mobility parameter $\epsilon$ needs to satisfy the following equation:}
\begin{equation}
    a_0(1 - \epsilon) + a_1 + \dots + a_{\tau} = s_{1Y}.
\end{equation}
{The solution can be found numerically for any value of $\tau$ but case $\tau=1$ admits an analytical expression for $\epsilon$:}
\begin{equation}
    \epsilon = \frac{1}{s_{1Y}} - 1.
\end{equation}

{The model dynamics is now described by an extended transition matrix which describes both migration between suburbs and ``mobility states'' (from state ``migrated $i$ years ago'' to state ``migrated $i+1$ years ago'' or state 0 where mobility is unrestricted):}
\begin{equation}
    \Hat{P} = 
    \begin{pmatrix}
        (1-\epsilon) I & \epsilon H & 0 & 0 & \dots & 0\\
        0 & 0 & I & 0 & \dots & 0\\
        0 & 0 & 0 & I & \dots & 0\\
        \dots & \dots & \dots & \dots & \dots & \dots \\
        0 & 0 & 0 & 0 & \dots & I\\
        I & 0 & 0 & 0 & \dots & 0
    \end{pmatrix}.
\end{equation}
{Matrix $\Hat{P}$ contains matrices $\epsilon H$, $(1-\epsilon) I$ and identity matrix $I$ as its $N \times N$ blocks.
The transition matrix for a 5-year period can be calculated using the $5$-th power of $\hat{P}$. The share of stayers in each suburb $k$ can be calculated as follows:}
\begin{equation}\label{eq:migration_5Y_memory}
    s_{5Y} = \frac{\sum_{k=1}^{N} x_k \sum_{i=0}^{N} a_i  \sum_{j=0}^{N} \Hat{p}_{kk;ij}^{5}}{\sum_{k=1}^{N}x_k}
\end{equation}
{where $\Hat{p}_{kk;ij}^5$ is the $k$-th diagonal element in block $i,j$ of matrix $\Hat{P}^5$.}

\section{Additional figures}

\subsection{Independence of the equilibrium state on the relocation frequency}
\begin{figure}[h!]
\centering
    \includegraphics[height=3in]{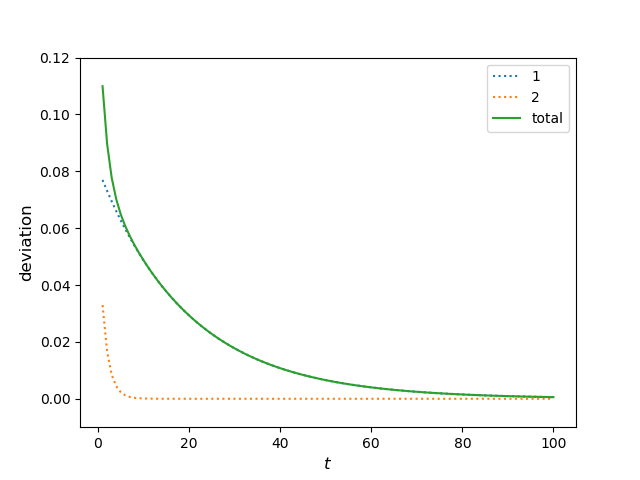}
    \caption{Convergence of $X_k(t)$ to the equilibrium $\alpha_k X_{eq}$ for $\alpha_1=0.7$, $\alpha_2=0.3$, $\epsilon_1=0.05$, $\epsilon_2=0.5$, matrix $H$ and initial conditions $X(0)$ are random, total population is 1. Equilibrium $X_{eq}$ is calculated as a left eigenvector of matrix $H$. Dotted lines 1 and 2 correspond to deviation $\Vert X_k(t) - \alpha_k X_{eq}\Vert$ as a function of time step $t$; solid line corresponds to deviation of the total structure $\Vert \sum_{k=1}^{C} X_k(t) - X_{eq}\Vert$.}
    \label{fig:convergence_example}
\end{figure}

Illustrated in \figr{fig:convergence_example}.

\subsection{Predicting number of movers with the 2011 data set}
\begin{figure}[h!]
\centering
    \includegraphics[height=7in]{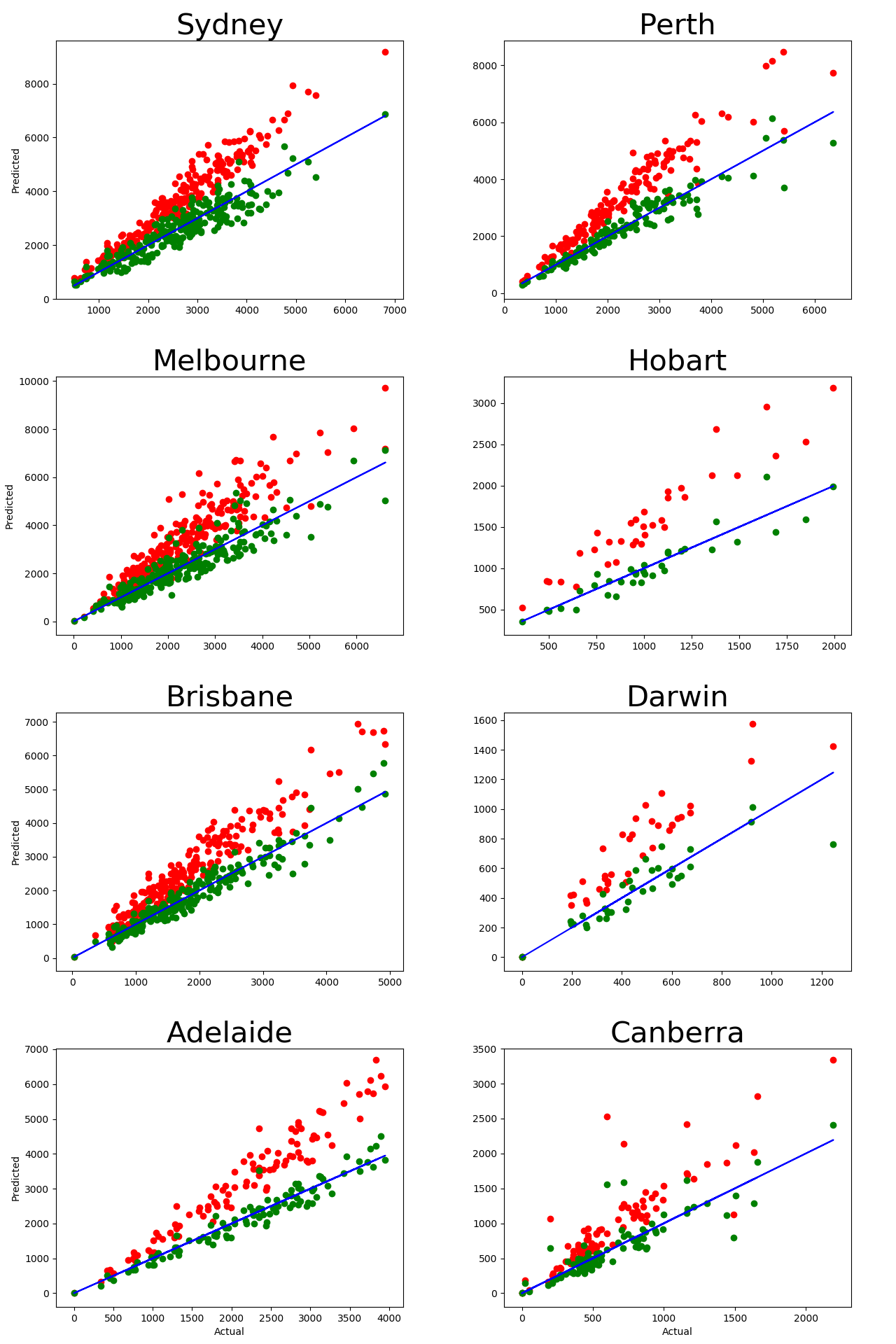}
    \caption{Number of movers in five-year migration data: actual ($\sum_{i \neq j}T_{ij}^{(5)}(2011)$) vs predicted by $P^5$ ($\sum_{i \neq j} \hat{T}_{ij}^{(5)}(2011)$), with each dot representing one suburb. Red dots correspond to the one-component model, the green dots correspond to the two-component model. The blue solid line has the slope of $1$, showing the ideal prediction. The corresponding calibration errors are shown in table~\ref{table:relative_error}.}
    \label{fig:migration_5Y_prediction_non_relocating_2011}
\end{figure}

Shown in \figr{fig:migration_5Y_prediction_non_relocating_2011}.

\medskip

\subsection{Actual population density map of the Australian capital cities}
\begin{figure}[h!]
\centering
    \includegraphics[height=7in]{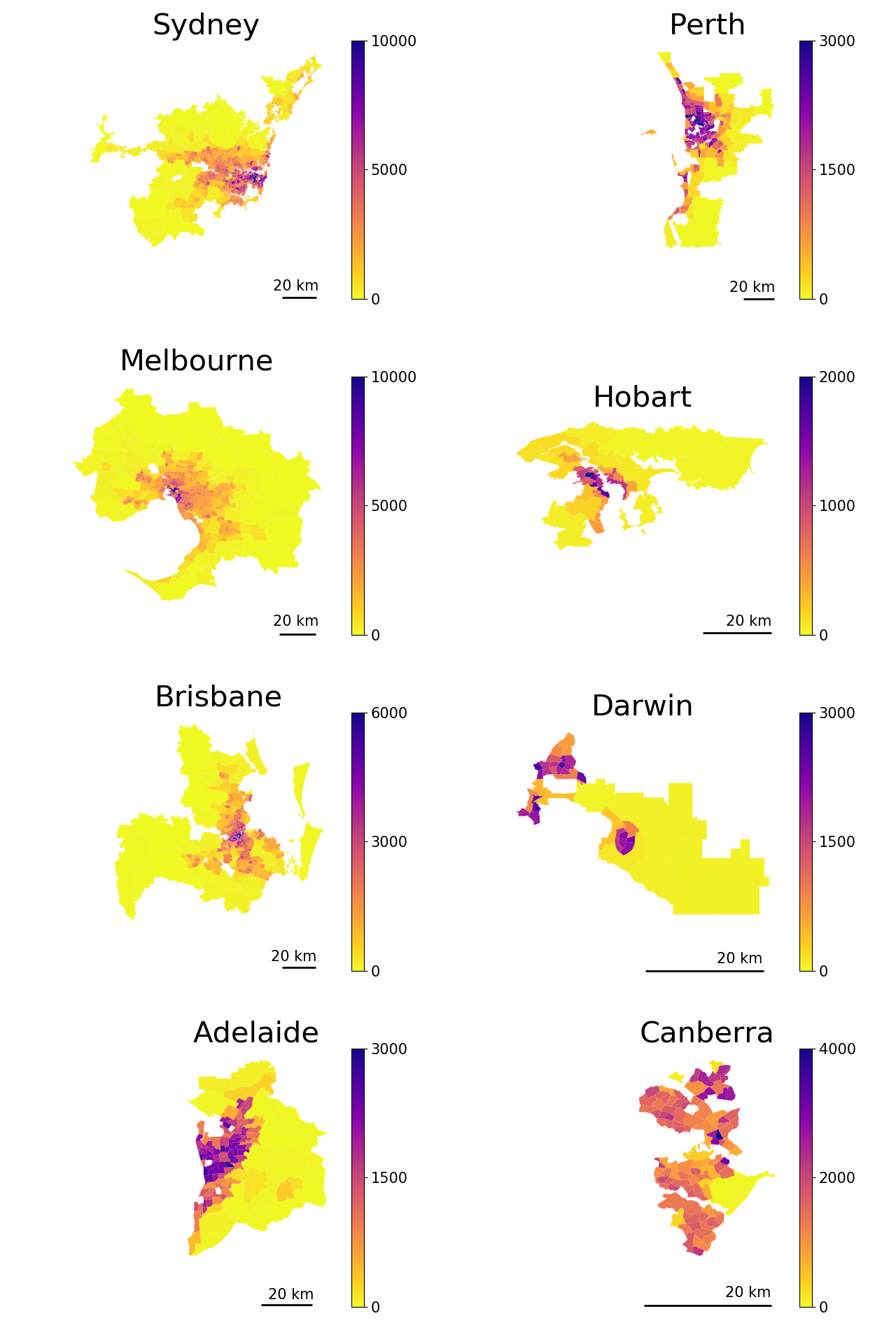}
    \caption{Actual population density map of the Australian capital cities (2016 Census). Scale bars in lower-left corners indicate distances equivalent to 20 km.}
    \label{fig:actual_population}
\end{figure}

Shown in \figr{fig:actual_population}.

\medskip

\subsection{Long-term prediction comparison: 2011 vs 2016}
\begin{figure}[h!]
\centering
    \includegraphics[height=7in]{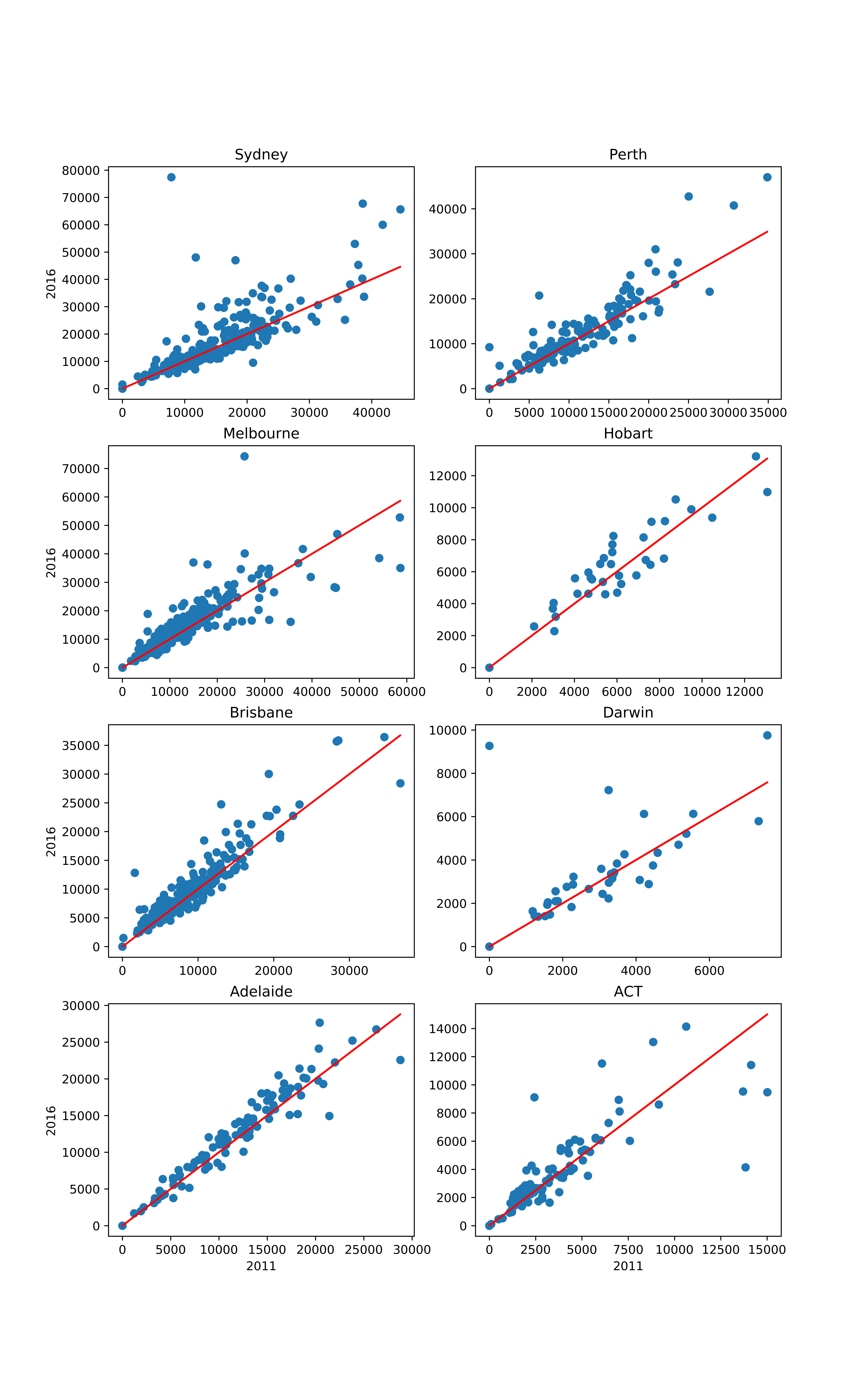}
    \caption{Equilibrium population of city suburbs predicted with 2016 data set is plotted against the 2011 prediction. The red solid line has the slope of $1$, showing the ideal consistency.}
    \label{fig:long_term_prediction_comparison}
\end{figure}

Shown in \figr{fig:long_term_prediction_comparison}.

\section{Additional tables}

\subsection{{Relative error of relocation prediction}}

{In this section, we provide relative errors for the five-year migration predictions shown in \figr{fig:migration_5Y_prediction_non_relocating} and \figr{fig:migration_5Y_prediction_non_relocating_2011}.
The errors are calculated as follows:}
\begin{equation*}
    e = \frac{\frac{1}{N}\sum_{i=1}^{N} |y_i^{predicted} - y_i^{actual}|}{\frac{1}{N}\sum_{i=1}^{N} y_i^{actual}},
\end{equation*}
{where $y_i^{actual}$ is actual migration outflow from suburb $i$ and $y_i^{predicted}$ is its predicted counterpart.
The corresponding values are shown in table~\ref{table:relative_error}.}

\begin{table}[h!]
\centering
    \caption{Relative error of the five-year migration prediction.}
    \vspace{3mm}
    \label{table:relative_error}
    \begin{tabular}{|| c || c | c || c | c || } 
        \hline
        \multirow{2}{*}{GCA} & \multicolumn{2}{c||}{2011}  & \multicolumn{2}{c||}{2016} \\
        \cline{2-5}
        & 1-component \!\! 
        & 2-component \!\! 
        & 1-component \!\! 
        & 2-component \\
        \hline\hline
        Sydney & 40\% & 12\% & 37\% & 13\% \\
        \hline
        Melbourne & 45\% & 14\% & 42\% & 11\% \\
        \hline
        Brisbane & 42\% & 10\% & 43\% & 10\% \\
        \hline
        Adelaide & 50\% & 9\% & 46\% & 7\% \\
        \hline
        Perth & 46\% & 9\% & 43\% & 10\% \\
        \hline
        Hobart & 54\% & 10\% & 48\% & 9\% \\
        \hline
        Darwin & 59\% & 15\% & 64\% & 16\% \\
        \hline
        Canberra & 51\% & 18\% & 40\% & 18\% \\
        \hline
    \end{tabular}
\end{table}

\subsection{{Five-year predictions of the single-component model}}\label{appendix:alternative_calibration}
{In this section, we compare five-year predictions of the single component model obtained for different ways of calibration. The share of stayers within a 5-year period is calculated based on the fifth power of one year migration rate matrix $P$. We use the following estimates for $P$:}
\begin{enumerate}
    \item {Estimated from the data directly (approach 1).}
    \item {Based on \eqref{eq:migration_matrix_decomposition}, with $\epsilon = 1 - s_{1Y}$ (approach 2).}
    \item {Based on \eqref{eq:migration_matrix_decomposition}, with $\epsilon = \frac{1}{N}\sum_{i=1}^{N}p_{ii}$ (approach 3).}
\end{enumerate}

{The corresponding results are shown in table~\ref{tab:alternative_calibrations}. }

\begin{table}[h!]
    \centering
    \caption{Share of people who do not change their place of residence within 5-year period (actual vs predicted) based on 2011-2016 migration data.}\label{tab:alternative_calibrations}
    \begin{tabular}{| c || c | c | c | c |}
        \hline
        GCA & actual & approach 1 & approach 2 & approach 3 \\
        \hline \hline
        Sydney & 0.733 & 0.630 & 0.624 & 0.630 \\
        \hline
        Melbourne & 0.734 & 0.618 & 0.610 & 0.613 \\
        \hline
        Brisbane & 0.702 & 0.569 & 0.560 & 0.545 \\
        \hline
        Adelaide & 0.752 & 0.635 & 0.632 & 0.642 \\
        \hline
        Perth & 0.709 & 0.584 & 0.579 & 0.588 \\
        \hline
        Hobart & 0.783 & 0.678 & 0.674 & 0.669 \\
        \hline
        Darwin & 0.713 & 0.527 & 0.518 & 0.511 \\
        \hline
        Canberra & 0.720 & 0.615 & 0.605 & 0.585 \\
        \hline
    \end{tabular}
\end{table}

\medskip

\end{document}